\documentclass[authoryear,preprint,12pt]{elsarticle}

\usepackage{amssymb}
\usepackage{subfig,graphicx}%,showframe}
\usepackage{color,soul}
\usepackage{setspace} 
\usepackage{todonotes}
\usepackage{appendix}
\usepackage[T1]{fontenc}

\journal{arXiv}

\begin{document}

\begin{frontmatter}



\title{Coastal Water Quality Prediction Based on Machine Learning with Feature Interpretation and Spatio--temporal Analysis}
%\todo{ Spatio--temporal malo je rubno al oej}

\author[inst1,inst3]{Luka Grbčić}
\ead{lgrbcic@riteh.hr}
\author[inst1,inst3]{Siniša Družeta}
\ead{sinisa.druzeta@riteh.hr}
\author[inst2,inst3]{Goran Mauša}
\ead{gmausa@riteh.hr}
\author[inst8]{Tomislav Lipić}
\ead{tlipic@irb.hr}
\author[inst3,inst5,inst6]{Darija Vukić Lušić}
\ead{darija.vukic.lusic@uniri.hr}
\author[inst1]{Marta Alvir}
\ead{malvir@riteh.hr}
\author[inst1,inst3]{Ivana Lučin}
\ead{ilucin@riteh.hr}
\author[inst1,inst3]{Ante Sikirica}
\ead{ante.sikirica@uniri.hr}
\author[inst9]{Davor Davidović}
\ead{davor.davidovic@irb.hr}
\author[inst3,inst4]{Vanja Travaš}
\ead{vanja.travas@uniri.hr}
\author[inst3,inst7]{Daniela Kalafatović}
\ead{daniela.kalafatovic@biotech.uniri.hr}
\author[inst10]{Kristina Pikelj}
\ead{kpikelj@geol.pmf.hr}
\author[inst10]{Hana Fajković}
\ead{hanaf@geol.pmf.hr}
\author[inst1]{Toni Holjević}
\ead{tholjevic@riteh.hr}
\author[inst1,inst3]{Lado Kranjčević\corref{cor1}}
\cortext[cor1]{Corresponding author}
\ead{lado.kranjcevic@riteh.hr}

\affiliation[inst1]{organization={Department of Fluid Mechanics and Computational Engineering, Faculty of Engineering, University of Rijeka},
            addressline={Vukovarska 58}, 
            city={Rijeka},
            postcode={51000},
            country={Croatia}}

\affiliation[inst2]{organization={Department of Computer Engineering, Faculty of Engineering, University of Rijeka},
            addressline={Vukovarska 58}, 
            city={Rijeka},
            postcode={51000}, 
            country={Croatia}}

\affiliation[inst3]{organization={Center for Advanced Computing and Modelling, University of Rijeka},
            addressline={Radmile Matejčić 2}, 
            city={Rijeka},
            postcode={51000}, 
            country={Croatia}}

\affiliation[inst4]{organization={Department of Hydrotechnics and Geotechnics, Faculty of Civil Engineering, University of Rijeka},
            addressline={Radmile Matejčić 3}, 
            city={Rijeka},
            postcode={51000}, 
            country={Croatia}}

\affiliation[inst5]{organization={Department of Environmental Health, Faculty of Medicine, University of Rijeka},
            addressline={Braće Branchetta 20/1}, 
            city={Rijeka},
            postcode={51000}, 
            country={Croatia}}

\affiliation[inst6]{organization={Department of Environmental Health, Teaching Institute of Public Health of Primorje-Gorski Kotar County},
            addressline={Krešimirova 52a}, 
            city={Rijeka},
            postcode={51000}, 
            country={Croatia}}
            
\affiliation[inst7]{organization={Department of Biotechnology, University of Rijeka},
            addressline={Radmile Matejčić 2}, 
            city={Rijeka},
            postcode={51000}, 
            country={Croatia}}
            
\affiliation[inst8]{organization={Laboratory for Machine Learning and Knowledge Representations, Rudjer Boskovic Institute},
	addressline={Biljenička cesta 54}, 
	city={Zagreb},
	postcode={10000},
	country={Croatia}}
            
\affiliation[inst9]{organization={Center for Informatics and Computing, Ruđer Bošković Institute},
            addressline={Biljenička cesta 54}, 
            city={Zagreb},
            postcode={10000}, 
            country={Croatia}}          

\affiliation[inst10]{organization={Department of Geology, Faculty of Science, University of Zagreb},
            addressline={Horvatovac 102a},
            city={Zagreb},
            postcode={10000}, 
            country={Croatia}}   
   
%SUGESIJA ZE ABSTRACT:
%Uvesti FIB???
%vidi  na verzije receniciza 

\begin{abstract}
Coastal water quality management is a public health concern, as poor coastal water quality can potentially harbor pathogens that are dangerous to human health. Tourism-oriented countries need to actively monitor the condition of coastal water at tourist popular sites during the summer season.
In this study, routine monitoring data of $Escherichia\ Coli$ and enterococci across 15 public beaches in the city of Rijeka, Croatia, were used to build machine learning models for predicting their levels based on environmental parameters as well as to investigate their dynamics and relationships with environmental stressors. Gradient Boosting algorithms (Catboost, Xgboost), Random Forests, Support Vector Regression and Artificial Neural Networks were trained with routine monitoring measurements from all sampling sites and used to predict $E.\ Coli$ and enterococci values based on environmental features. 
The evaluation of stability and generalizability with 10-fold cross validation analysis of the machine learning models, showed that the Catboost algorithm performed best with R$^2$ values of 0.71 and 0.68 for predicting $E.\ Coli$ and enterococci, respectively, compared to other evaluated ML algorithms including Xgboost, Random Forests, Support Vector Regression and Artificial Neural Networks.
We also use the SHapley Additive exPlanations technique to identify and interpret which features have the most predictive power. The results show that site salinity measured is the most important feature for forecasting both $E.\ Coli$ and enterococci levels.
Finally, the spatial and temporal predictive accuracy of both ML models were examined at sites with the historically lowest coastal water quality. The spatial $E. Coli$ and enterococci models achieved strong R$^2$ values of 0.85 and 0.83, while the temporal models achieved R$^2$ values of 0.74 and 0.67. The temporal model also achieved moderate R$^2$ values of 0.44 and 0.46 at a site with historically high coastal water quality.

\end{abstract}



\begin{keyword}
coastal water quality \sep machine learning  \sep shap \sep catboost \sep fecal indicator bacteria 
\end{keyword}

\end{frontmatter}

%\linenumbers
\doublespacing 
\section{Introduction}

Predicting the coastal water quality at public beaches would provide great benefits for the general human population as it is a major health issue due to a potential contamination with pathogenic microorganisms. 

Currently, in countries with a strong tourism sector, outdated information on coastal water quality is given because on-site measurements and laboratory analyses are time-consuming and lag behind the time frame in which a warning to the public would be necessary. For example, in Croatia, a country dependent on tourism in summer, sampling and laboratory analysis takes on average 2.2 days \citep{luvsic2017temporal}.

Furthermore, a lack of consistent criteria for estimating the potentially hazardous level of bacterial pollution is evident in the regulations between the EU and the US, with the EU directives using \textit{Escherichia Coli} (EC) and enterococci (ENT) which are considered Faecal Indicator Bacteria (FIB) \citep{luvsic2017temporal,directive2006directive}, while the US regulations only consider ENT as the main indicator \citep{usepa2012recreational}.

To accelerate the process of determining the safety of public beaches, Machine learning (ML) based data-driven methods have been increasingly used for the purpose of predictive modeling of FIB concentrations. The predictive modeling of FIB concentrations aims to solve the problem of the slow measurement-based issuance process that a certain location is unsuitable for bathing. 

However, due to the great complexity of the dynamics of FIB, which involves an interaction of many environmental parameters, the application of ML based approaches require much further research to increase their predictive power.

In previous studies, the Random Forest (RF) algorithm was used to predict FIB concentrations at 5 different beaches and showed that a logarithmic transformation of raw measurements increases the prediction accuracy of the ML model \citep{parkhurst2005indicator}. Also, Tree Regression (TR) and RF models were used to predict FIB values in freshwater, and it was found that precipitation is an important parameter for FIB prediction \citep{jones2013hydrometeorological}.

Five different algorithms for FIB prediction on a single location using data collected 5 times a week for six bathing seasons were compared \citep{thoe2014predicting}. The trained models were auto-regressive and it was observed that the Artificial Neural Network (ANN) model performs good in terms of predicting the days when FIB exceeded levels of sufficient water quality. 

Gradient Boosting (GB) produces good results for FIB prediction at 7 different lake beaches \citep{brooks2016predicting}. The used data was collected for 3 bathing seasons, 2 to 4 times per week for 12-14 weeks per season.

A hybrid wavelet auto-regressive ANN  (WA-NAR) was used to predict EC concentrations at four different lake beaches and achieved high accuracy \citep{zhang2018real}. The study increased the EC measured data temporal resolution with the Monte Carlo Markov Chain (MCMC) method in order to increase the training and testing sets for the ML model. 

It was found that the ANN model performs better for FIB prediction than SVR for two beaches \citep{park2018development}. The studied sampling sites have a proximity to a waster water plant and a waste effluent was considered as a feature. 

A stacking ML model was created to predict FIB values at 3 lake beaches \citep{wang2021improving}. It was concluded in the study that an ensemble of several ML algorithms improves the overall accuracy of FIB prediction.

While most previous research focused on ML regression models which try to predict the exact value of FIB, in the recent work by \citep{xu2020predictive}, a classification approach was tested ML algorithms in conjunction an oversampling method ADASYN in order to balance the classification.

Recently, Bayesian Belief Networks (BBN) have also been used to classify whether FIB concentration values exceed a certain predefined threshold \citep{avila2018evaluating,panidhapu2020integration, yuniarti2021application}.

RF model achieved high daily FIB predictive accuracy (R$^2$ value of 0.95) \citep{searcy2021day} using FIB data obtained with high-frequency sampling at only 10-minute measurement intervals at three different beaches.

Predictive FIB modeling includes both temporal and spatial ML models. A temporal FIB ML model is used to predict future FIB values at specific locations based on measured historic data, while a spatial ML model is used to assess locations where no FIB measurements exist.

In this work, a spatial and temporal ML prediction of FIB concentrations was investigated with data collected at 14 different locations through the national routine FIB monitoring program. 
The majority of the sampling locations have a historically high quality of coastal water, with the exception of three locations at the easternmost part of the cluster which are in the focus of the ML analysis, hence, the main goal of the study was to assess the accuracy of the spatial and temporal ML prediction on three of the most problematic locations in the studied cluster of public beaches in the city of Rijeka, Croatia.

Additionally, we investigate the dynamics of both EC and ENT bacteria with ML methods at the cluster of beaches.
ML algorithms such as Gradient Boosting (Catboost, Xgboost), RF, SVR and ANN were used to train models with routine monitoring data collected at all locations and separate models were created for EC and ENT measurements. A variety of environmental features were used as ML model inputs, and those include on site measurements (temperature, salinity) and other bulk meteorological features which exhibit correlation with FIB.

Furthermore, the SHapley Additive exPlanations (SHAP) method for ML model interpretability was used to show which features are the greatest drivers of FIB dynamics overall at the studied cluster of beaches.

We show that a strong R$^2$ score is achievable with both spatial and temporal ML models on three of the most problematic locations in terms of coastal water quality. Additionally, with the SHAP method we identify that the feature with the strongest influence of the ML model output is salinity for both EC and ENT prediction.

\section{Materials and Methods}
\subsection{Sampling Locations and Data Acquisition}

\begin{figure}[!h]
\centering
\subfloat[]{\label{a}\includegraphics[width=.495\linewidth]{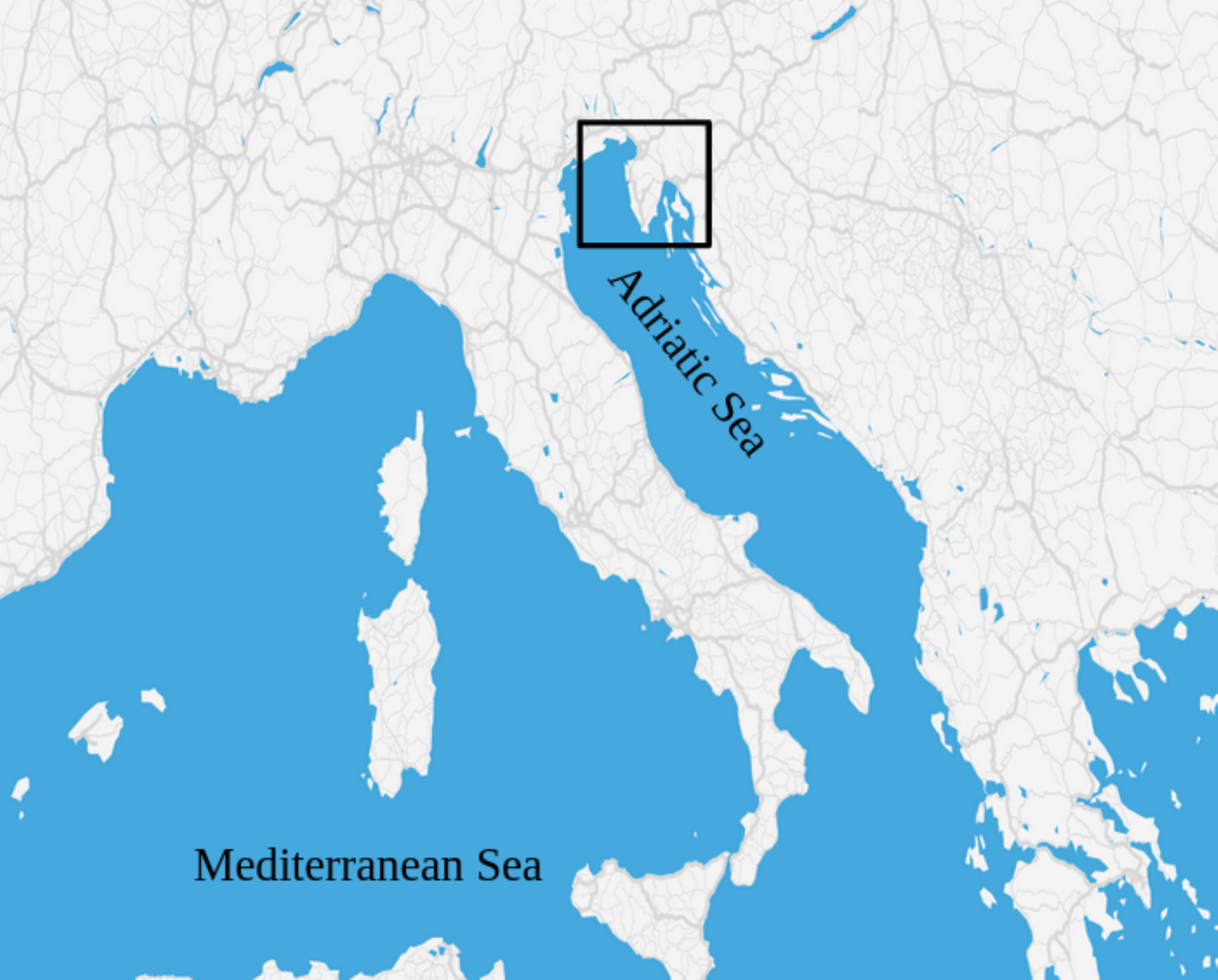}}
\hfill
\subfloat[]{\label{b}\includegraphics[width=.495\linewidth]{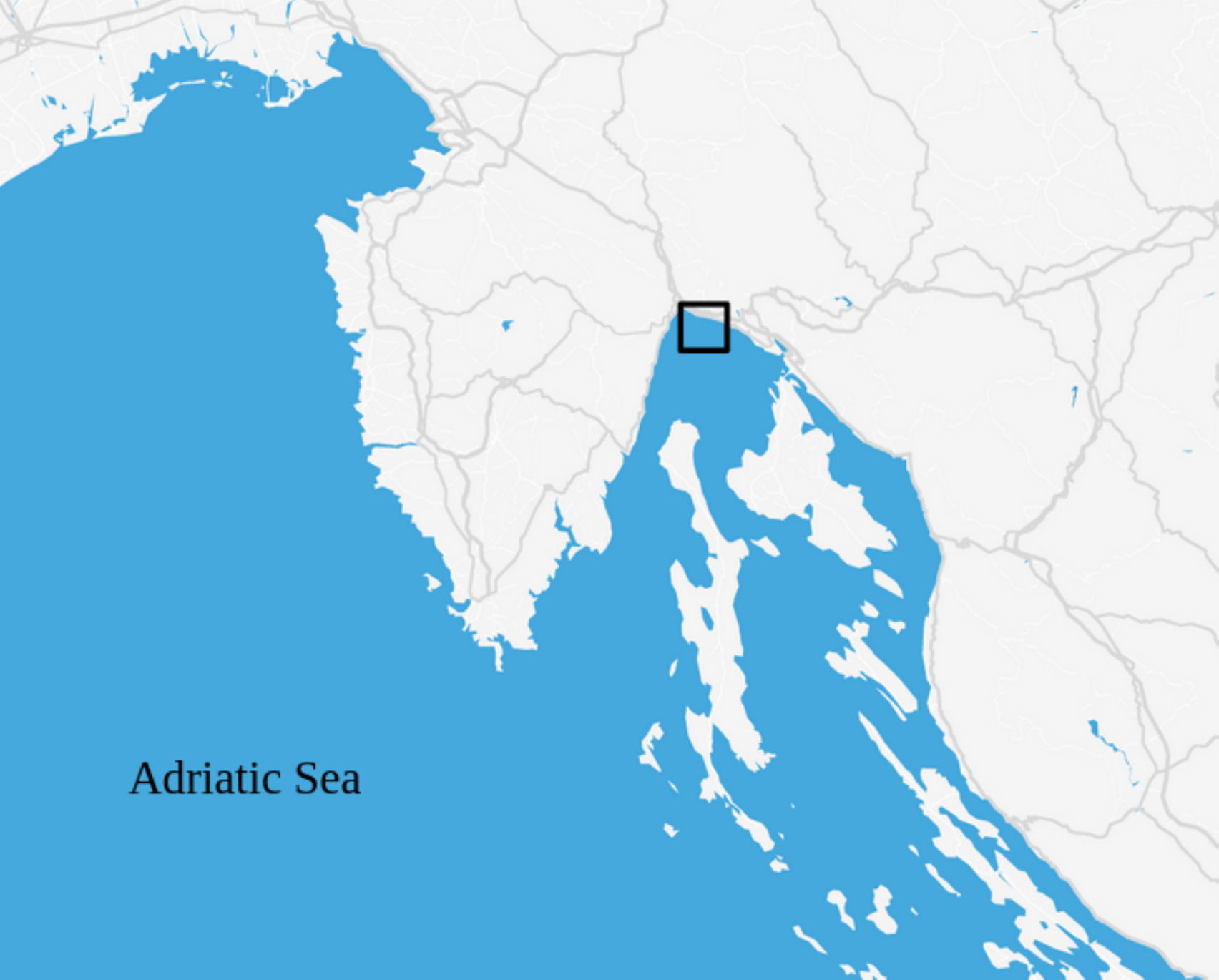}}
\par
\subfloat[]{\label{c}\includegraphics[width=\linewidth]{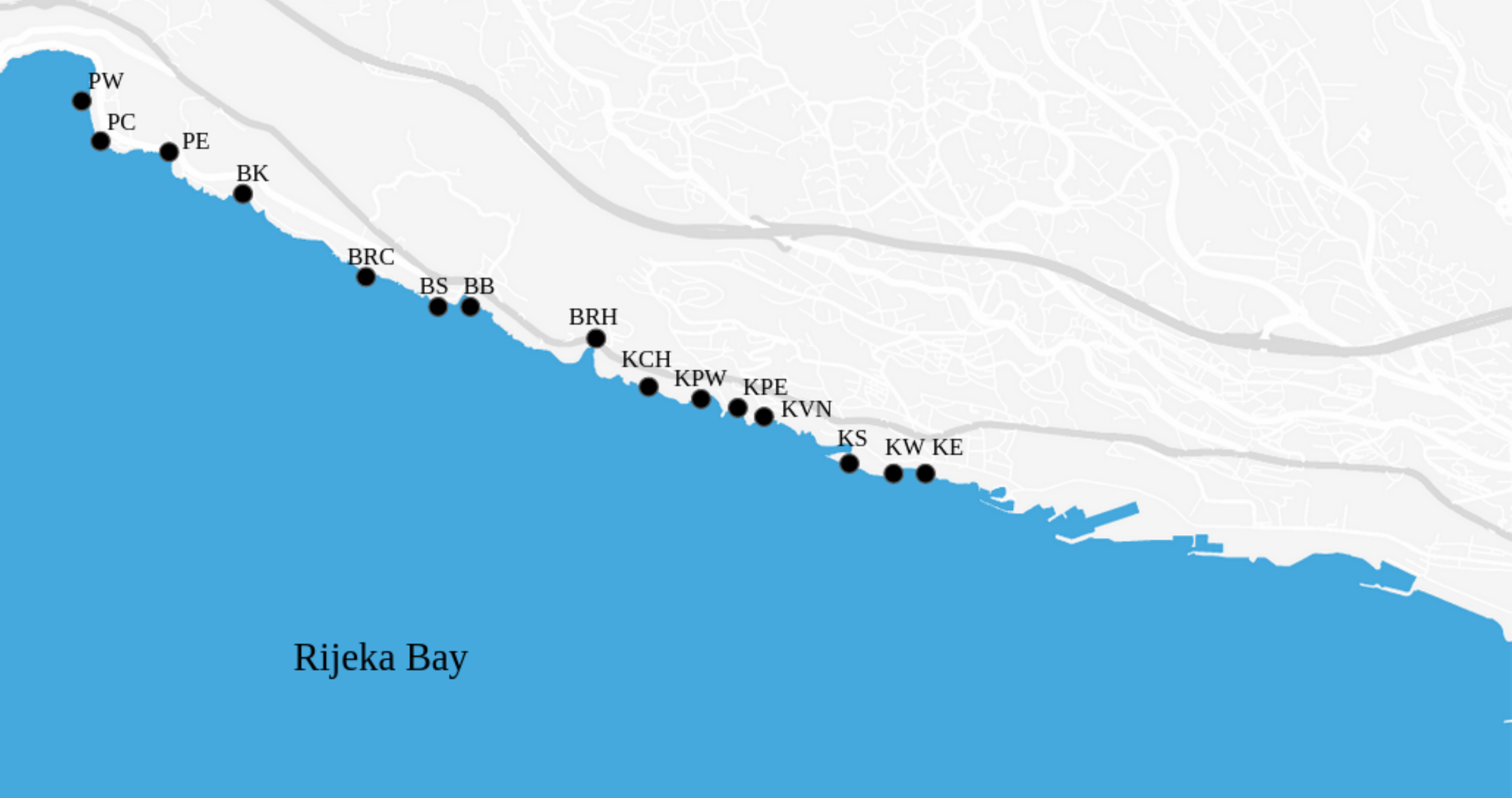}}
\caption{Location of Rijeka Bay and fourteen sampling sites used for data acquisition in order to create a ML model.}
\label{fig:sampling_loc}
\end{figure}

The predictive model was trained and tested on data obtained at 14 different locations at the western part of the city of Rijeka, Croatia, as presented in Fig. \ref{fig:sampling_loc}. Rijeka is located in the Primorje-Gorski Kotar County and has over 200000 inhabitants in its metro area, making it the third largest city in Croatia.  An overlapping between population density and coastal water quality was shown through spatial interpolation, further indicating that there exists an anthropic influence on coastal water quality \citep{malcangio2018statistical}.

Rijeka is also one of the wettest areas in Croatia with a recorded average annual rainfall of 1530 mm \citep{benac2003origine}. The connection between FIB concentrations and rainfall exists as rainfall is one of the most common input feature used for creating FIB ML and statistical prediction models \citep{luvsic2017temporal,parkhurst2005indicator,brion1999neural,he2019storm,ge2007some}.

Generally, the larger area of Rijeka is characterized by steep mountainous and karst terrain and it has been showed with previous extensive research which used tracer tests \citep{biondic1997hydrogeological, bonacci2018water} that its underground comprises a complex network of conduits, cavities and caves. The hydrogeological studies also indicate that the specific studied locations (beaches) also contain coastal karst springs and crevices which enable an infiltration of groundwater after heavy rainfall which occurs at locations at a higher altitude than the studied sites. 

\begin{figure}[h]
\centering
\includegraphics[width=\textwidth]{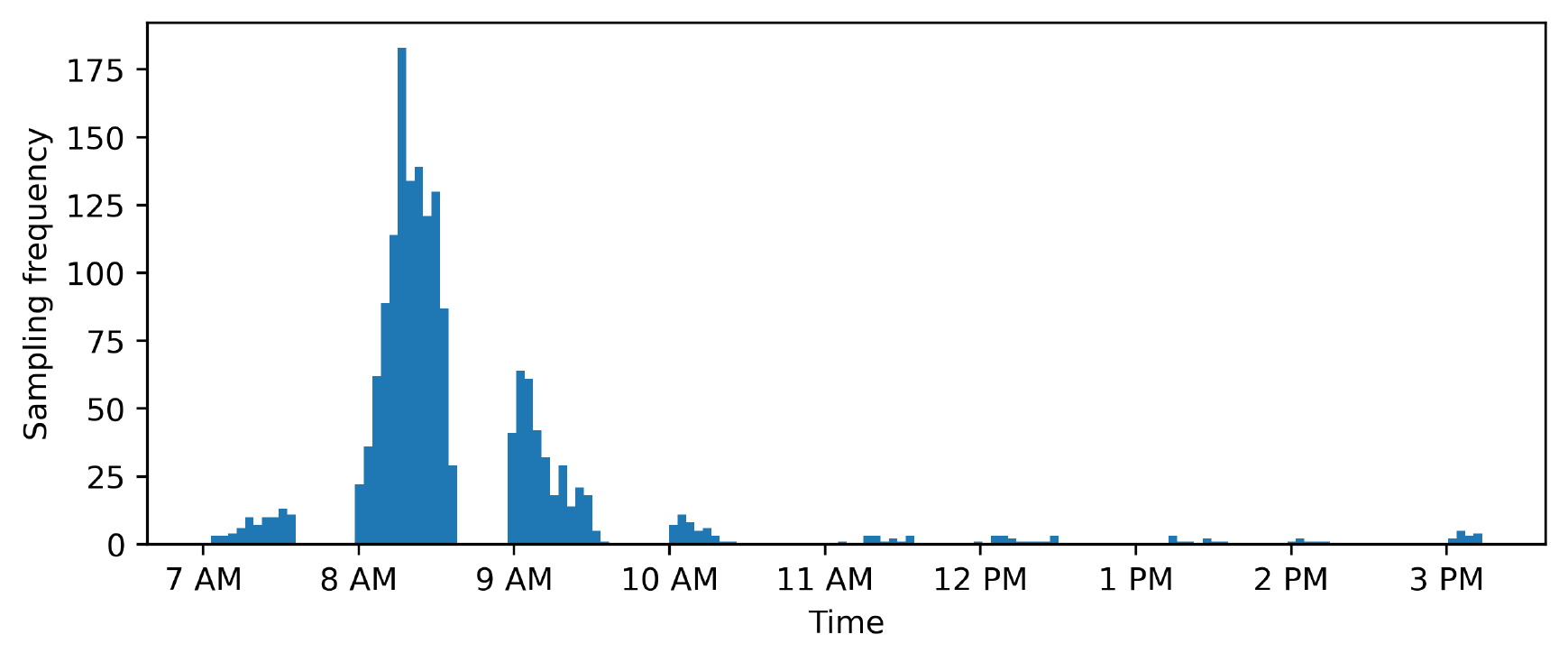}
\caption{EC and ENT sampling frequency distribution through a time period from 7 AM to 3 PM.}
\label{fig:sample_freq}
\end{figure}

The collected data included measurements of EC and ENT which are prescribed by the European Union Bathing Water Directive, 2006/7/EC (EU BWD) \citep{directive2006directive}.
Data which was used in this study is a part of the Croatian national routine monitoring of coastal water quality which includes a time period from 2009 to 2020 for all locations except for KPW (measurements started in 2010) and KS (measurements started in 2019). A total of 10 measurements were made at every location for each bathing season using a set of thoroughly described methods \citep{luvsic2017temporal}. Measurements were made every 15 days within the bathing season which started in mid-May and lasted until the end of September of each year and each measurement was made in the time span from 7 AM to 3 PM with most being between 8 AM and 9 AM. The sampling frequency distribution can be observed in Fig. \ref{fig:sample_freq}. Data obtained from all locations presented in Fig. \ref{fig:sampling_loc} were used to create a ML model. 

Statistical summary of the data for the three locations with the highest EC and ENT measurements (namely locations KE, KW, and KS) is given in Table \ref{tab:highest_stat}. All three locations are adjacent to each other in the eastern part of the studied sampling sites and are most problematic in terms of coastal water quality as they have the highest median and average values recorded through their respected time interval.

\begin{figure}[!h]
\centering
\includegraphics[width=\textwidth]{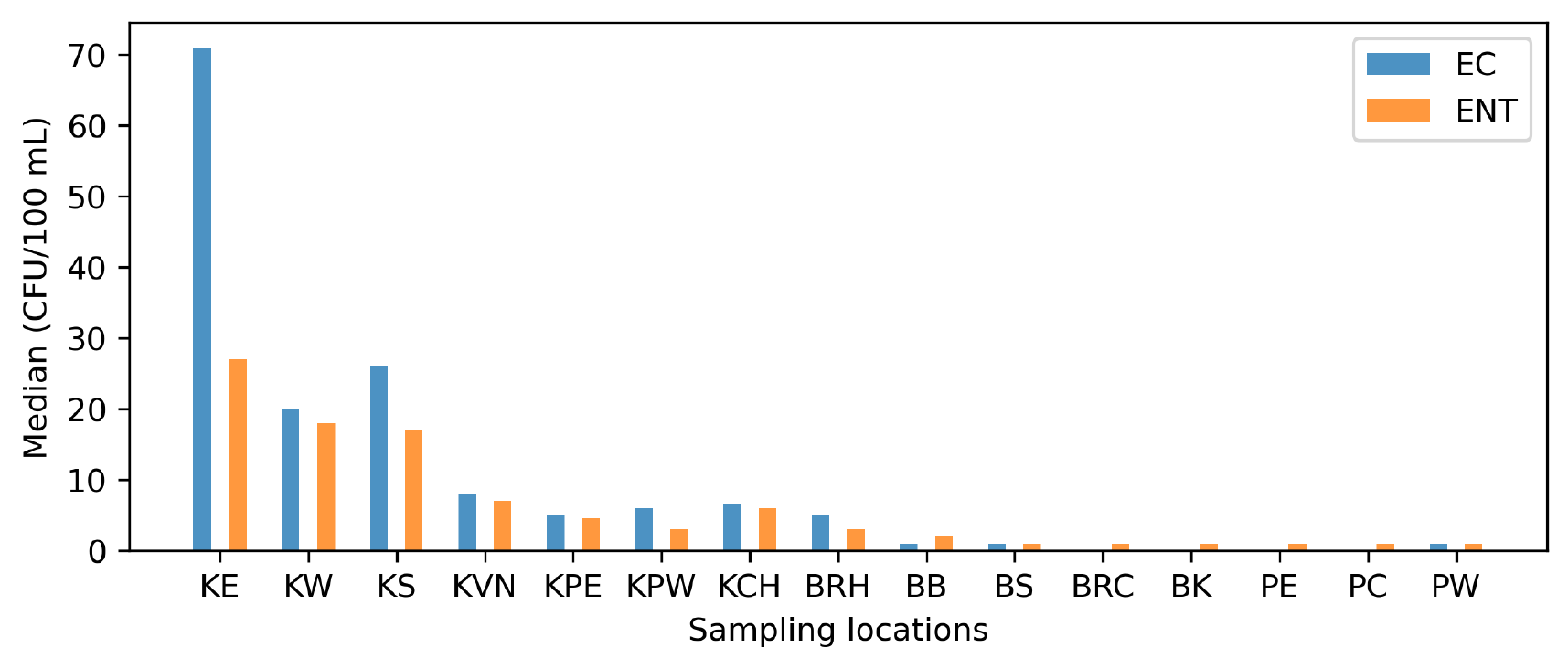}
\caption{Median values of EC and ENT at each sampling location. The x-axis represents the sampling locations with the sequence starting at the easternmost sampling site and ending at the westernmost sampling site. KS values account only for a time period from 2019 to 2020.}
\label{fig:median_ec_ent}
\end{figure}

Furthermore, Fig. \ref{fig:median_ec_ent} shows the median values of both EC and ENT in CFU/100 mL at all sampling locations of the studied cluster. The x-axis represents the sampling locations with the first one (KE) being the easternmost sampling location, while the last one (PW) is the westernmost sampling location. It can be observed that there exists a spatial trend in which the EC and ENT median values become lower from east to west.

\begin{table}[h]
\centering
\caption{Locations with the highest values of EC and ENT measurements and statistical parameters (CFU/100 mL). Med, avg, std dev, max represent the median, average, standard deviation and maximum values, respectively.}
\resizebox{\textwidth}{!}{%
\begin{tabular}{ c  c  c  c  c  c  c  c  c  c } 
\hline
\multicolumn{1}{ c }{\textbf{Location}}&  \multicolumn{1}{ c }{\textbf{Period}} & \multicolumn{4}{  c  }{\textbf{EC}} & \multicolumn{4}{  c  }{\textbf{ENT}}\\
\hline
 &   & \textbf{med} & \textbf{avg}  & \textbf{std dev} & \textbf{max} & \textbf{med} & \textbf{avg}  & \textbf{std dev} & \textbf{max}  \\
\hline
KE  &  2009--2020 & 72 & 115.9  & 267.8 & 800 & 27 & 44.5  & 80.3 & 190  \\
KW & 2009--2020 & 20 & 61.8  & 81.2 & 520 & 18 & 30.5  & 36 & 180 \\
KS & 2019--2020 & 26 & 56.8  & 78.1 & 290 & 17 & 35.5  & 48 & 180 \\
\hline
\end{tabular}}
\label{tab:highest_stat}
\end{table}

\newpage

The Croatian criteria for coastal water quality assessment differs from the EU criteria for the measured EC level, while for ENT it is the same \citep{gazette2008croatia}. 

In order for a beach to be considered safe for bathing activities, Croatian national rules prescribe that EC and ENT values must be below 300 CFU/100 mL and 185 CFU/100 mL, respectively, while the EU criteria defines values for EC of 500 CFU/100 mL for EC and for ENT of 185 CFU/100 mL, based upon a 90-percentile evaluation \citep{luvsic2017temporal}. The Croatian national criteria applies to as an assessment for every sample and as a yearly/final evaluation, while the EU applies only as yearly/final evaluation of coastal water quality.

If a beach is to be considered excellent by Croatian standards, an EC measurement must be below 150 CFU/100 mL and an ENT measurement must be below 100 CFU/mL. The excellent and good criteria requirements are defined by a 95-percentile confidence evaluation, while the sufficient criteria is based on a 90-percentile confidence evaluation.

\begin{figure}
  \centering
  \subfloat[a][]{\includegraphics[width=\textwidth]{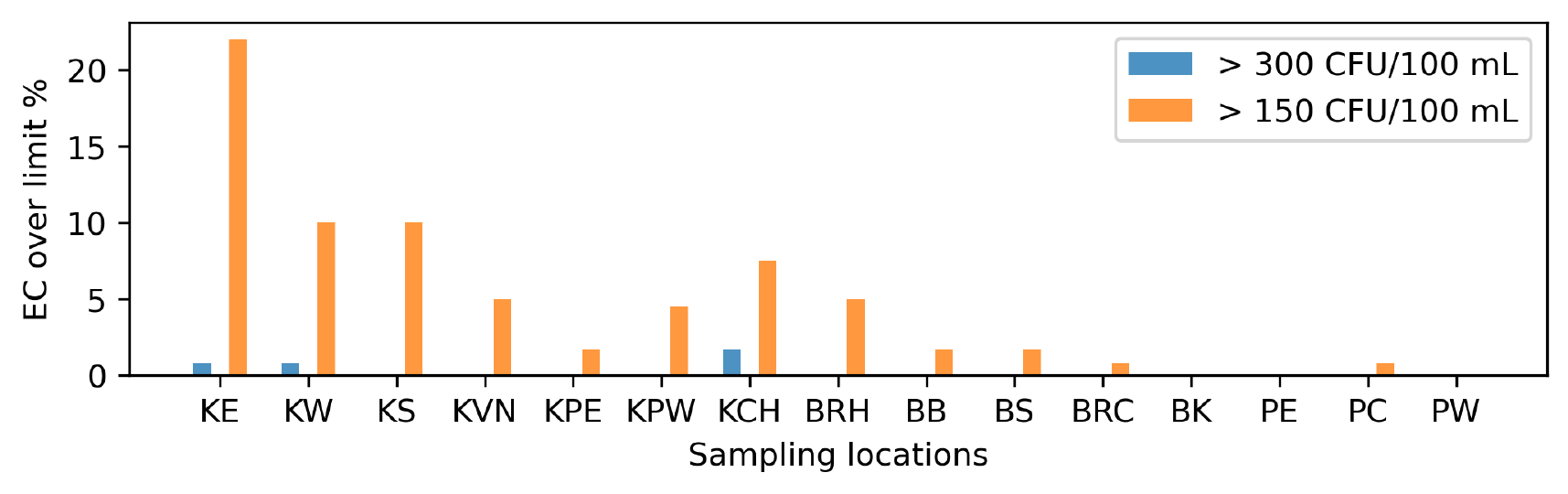} \label{fig:ec_locations}} \\
  \subfloat[b][]{\includegraphics[width=\textwidth]{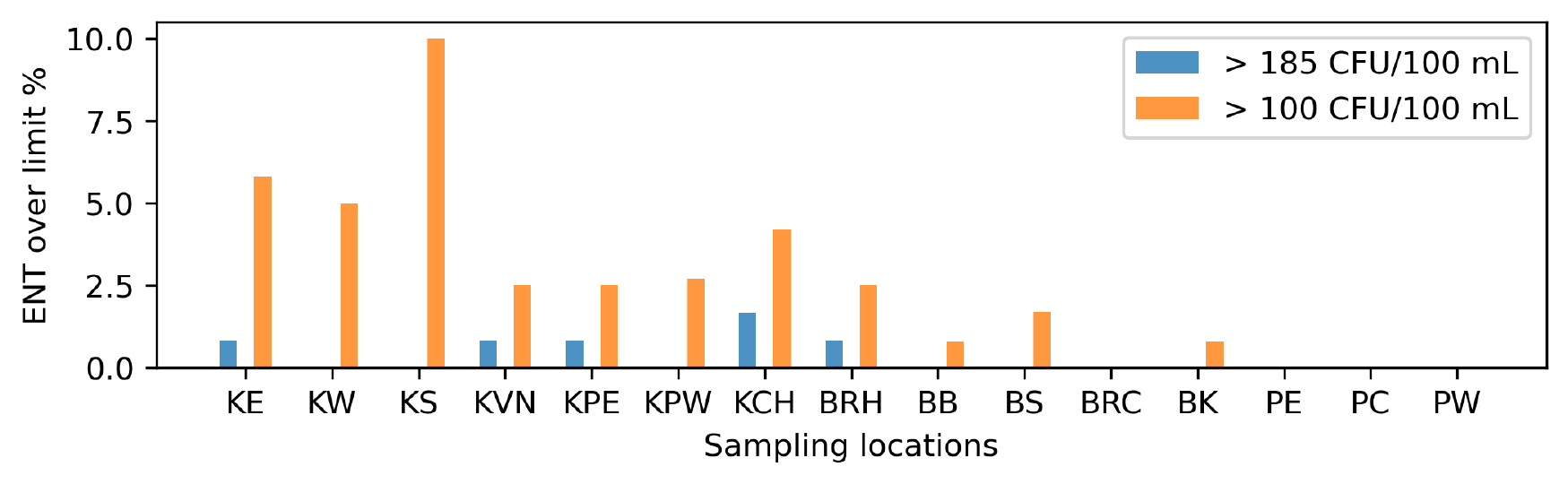} \label{fig:ent_locations}}
  \caption{Percentage of measurements at each sampling location when FIB values exceeded the Croatian criteria for coastal water quality both for being excellent and sufficiently safe. Sampling location KS accounts for a time interval from 2019 to 2020, KPE for 2010 to 2020, while all other locations for 2009-2020. The y-axis represents the over limit percentages for both EC and ENT.} 
  \label{fig:ec_ent_criteria} 
\end{figure}

In Fig. \ref{fig:ec_locations}, the percentage of EC measurements when a sampling location was classified as sufficiently safe and excellent can be observed, while Fig. \ref{fig:ent_locations} provides the same data for ENT measurements. A total of 120 measurements were made at every sampling location except for KPW where 110 measurements were made, and KS where the total number was 20. It is noticeable that the eastern part of the studied cluster, which includes locations KE, KW and KS, is the most problematic in terms of coastal water quality. However, with a total of 1690 measurements, it can also be concluded that the event of exceeding the criteria of being sufficiently safe for both EC and ENT is extremely rare at every sampling location.

\newpage
\subsection{Environmental Parameters}

Environmental parameters used as ML model input features included physical-chemical properties of the environment where and when the sample was taken and meteorological properties both at the time of sample acquisition and their antecedent cumulative values. A total of 33 features were considered for predictive modelling and are listed in Table \ref{tab:env_parameters}. 

\begin{table}[!h]
\caption{Environmental features used for building a ML predictive model. Antecedent Cumulative Precipitation and GHI account for 8 features each, while Antecedent GHI accounts for 4 additional features.}
\centering
\resizebox{\textwidth}{!}{
\begin{tabular}{ccc}
\hline
\textbf{Feature}	&\textbf{Unit}	& \textbf{Source}  \\ 
\hline
Air temperature ($T_a$) & $^{\circ}$C &  \textit{In situ} \\
Salinity ($S$)& - & \textit{In situ}  \\
Sea temperature	($T_s$)& $^{\circ}$C &  \textit{In situ} \\
Water level ($WL$)& m & IZOR$^*$ \\ 
Antecedent Cumulative Precipitation$^{\dagger}$ ($CPrec$) & mm & DHMZ$^{\ddagger}$ \\
Global Horizontal Irradiance (GHI) & W/m$^2$ &  DHMZ \\ 
Antecedent Cumulative GHI$^{\dagger}$ ($CGHI$) & W/m$^2$  &  Solcast \\
Antecedent GHI$^{\mathsection}$ ($GHI_i$)& W/m$^2$  &  Solcast \\
Dewpoint temperature ($T_d$)& $^{\circ}$C  & Solcast \\
Precipitable water ($PW$) & kg/m$^2$  &  Solcast \\
Relative humidity ($RH$) & \%  &  Solcast \\
Surface Pressure ($SP$) & hPa  &  Solcast \\
Wind speed ($WS$) & m/s  & Solcast \\
Wind direction ($WD$) & $^{\circ}$  & Solcast \\
\hline
\multicolumn{1}{l}{$^*$Institute of Oceanography and Fisheries, Split, Croatia}
\\
\multicolumn{1}{l}{$^{\dagger}$Features for antecedent periods of 4 hours and 2, 3, 4, 7, 14, 30 and 60 days.} \\
\multicolumn{1}{l}{$^{\ddagger}$Croatian Meteorological and Hydrological Service}
\\
\multicolumn{1}{l}{$^{\mathsection}$Features for antecedent periods of 1, 2, 3 and 4 hours.} \\
\label{tab:env_parameters}
\end{tabular}}
\end{table}

Salinity, sea and air temperature were all measured at the sampling locations at the same time the EC and ENT data was collected. Water level data was obtained from the Institute of Oceanography and Fisheries (IZOR), Split, Croatia \citep{IZOR} and it is a prediction generated by the XTide software \citep{flater1996brief}. Water levels were linearly interpolated between two predicted water level points to exactly fit the time of sampling at each location. 

The precipitation data was obtained from the Croatian Meteorological and Hydrological Service (DHMZ) and the values were reported for every hour of the day. The precipitation data is used to calculate the cumulative values for antecedent intervals of 4 hours, and 2, 3, 4, 7, 14, 30 and 60 days before the exact time the sampling was made at each location. A total of 8 different features of precipitation data were created with the antecedent intervals for the ML model. The antecedent precipitation features are described in subsection \ref{sub:mlr}.

All other features were obtained at hourly intervals from Solcast, which is the solar resource assessment and forecasting data enterprise \citep{Solcast}. Solcast data has previously been validated and recommended for research purposes in the work by \citep{bright2019solcast}. All Solcast features were also linearly interpolated to match the sampling times. It should be noted that the air dewpoint temperature and relative humidity are recorded at 2 meters above ground level, the surface pressure accounts for the atmospheric pressure at ground level, while both wind speed and direction are recorded at 10 meters above ground level. 

The antecedent cumulative global horizontal irradiance (GHI) is calculated the same way and for the same antecedent intervals as cumulative precipitation, creating 8 additional features. Lastly, the values of GHI recorded 1, 2, 3 and 4 hours before the EC and ENT samples were collected were also considered as features (antecedent GHI). The created cumulative GHI and antecedent GHI features are described in subsection \ref{sub:mlr}.

\newpage

\subsection{Predictive Modeling}
\subsubsection{ML Regression}
\label{sub:mlr}

The input features used to train and test the regression ML model are features listed in Table \ref{tab:env_parameters}, while the outputs of the ML model are the measured FIB values which correspond to each instance or state of the input features. 

\begin{figure}[!h]
\centering
\includegraphics[width=\textwidth]{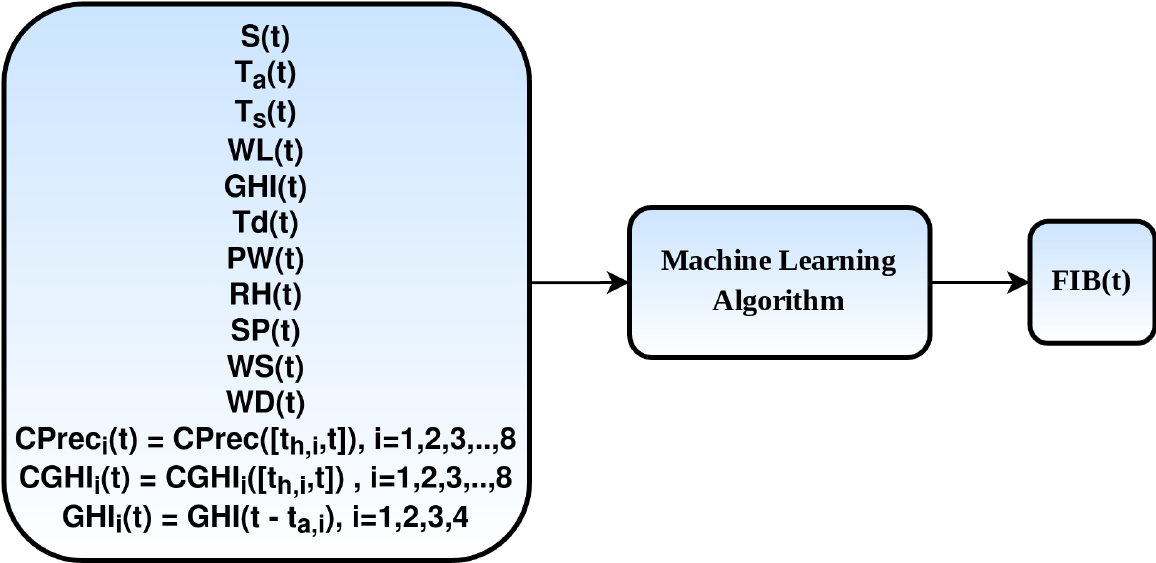}
\caption{The ML regression model flow chart. Environmental features are used in conjunction with a ML algorithm in order to train a ML model which predicts the FIB values.}
\label{fig:flowchart}
\end{figure}

In Fig. \ref{fig:flowchart} the flowchart of the ML model is presented. Most features represent values at measurement time $t$, while CPrec and CGHI correspond to the antecedent intervals from measurement time $t$ to $t_h$, where $t_h \in$ \{{4 hours, 2, 3, 4, 7, 14, 30, 60 days}\}, and the values of $GHI_i$ correspond to times $t_a$ (where $t_a \in$ \{{1, 2, 3, 4 hours}\}) before the measurement time $t$.

\subsubsection{ML Algorithms}

The ML algorithms used for EC and ENT values prediction were Gradient Boosting (Catboost, Xgboost), Random Forests, Support Vector Regression and Artificial Neural Networks (Multilayer Perceptron). 

Catboost (CB) is a gradient boosting toolkit which includes algorithmic advances such as the implementation of ordered boosting, a permutation-driven alternative to the classic gradient boosting algorithm, and an algorithm categorical features processing \citep{dorogush2018catboost}. The algorithm improvements made in Catboost are created to solve the prediction shift problem caused by target leakage.

Xgboost (XGB) is a highly scalable tree boosting framework which incorporates a sparsity-aware algorithm and weighted quantile sketch for approximate tree learning \citep{chen2016xgboost}. Xgboost combines cache access patterns, high data compression and sharding in order to build a tree boosting system.

Random Forests (RF) is an ensemble ML algorithm which creatures multiple decision trees defined with features selected at random \citep{breiman2001random}. The decision trees with random features have both a lower variance and are less likely to cause overfitting of the ML model.

A Multilayer Perceptron (MLP) Artificial Neural Network (ANN) consists of three basic layers (input, hidden and output) of artificial neuron nodes. Multiple hidden layers could be used in the MLP and each neuron in the hidden and output layers uses a nonlinear activation function which mimics the workings of human a brain \citep{nielsen2015neural}.

Support Vector Machines (SVM) are a class of supervised ML algorithms in which a superposition of kernel functions is used for data approximation \citep{drucker1997support}. Support Vector Regression (SVR) is a variation of SVM which is specifically used for to train ML regression models.

Python 3.8. implementation of all algorithms was used in this study and Table \ref{tab:ml_algorithms} lists all of the algorithms used to train spatial and temporal FIB models. CB and XGB algorithms haven't been previously used to train FIB prediction models.

All EC and ENT measurements were log$_{10}$ transformed before ML model training as it was shown in previous studies that this procedure reduces both the variability and the skewness of EC and ENT values \citep{parkhurst2005indicator,searcy2021day,thoe2015sunny, searcy2018implementation}, and it proved to be beneficial for the regression ML in this study.

Before SVR and ANN algorithms were used for model training and testing, all input features were standardized with the $StandardScaler$ function of the scikit-learn 0.24.1, Python 3.8. module. The multivariate input data is transformed in a way that each feature is independently standardized, more specifically, the mean value of each feature is subtracted from all instances and divided by the standard deviation of the same feature. This process has shown to be beneficial for the SVR and ANN algorithms, however, it did not affect CB, XGB and RF, and hence the input data remained unchanged when used for model training of the three algorithms.

The number of estimators for CB, XGB and RF was set to be 550, and all other hyper-parameters were set as default values. The radial basis function kernel was used with the epsilon-SVR algorithm, and both the epsilon-band $\epsilon$ and penalty C hyperparameter values were optimized using the Fireworks Algorithm (FWA) \citep{tan2010fireworks} within the Python 3.8. numerical optimization module indago 0.1.9 \citep{druzeta2020}.

The ANN consisted of two hidden layers which consisted of 100 neurons each, the rectified linear unit activation function for the hidden layer and the stochastic gradient-based optimizer \emph{adam} was used for model optimization. The number of layers and neurons of the ANN was also optimized with FWA. All other SVR and ANN hyper-parameters were set as default values.

\begin{table}[!h]
\caption{ML algorithms used for EC and ENT predictive modeling. The FIB reference column lists recent FIB predictive modeling research which used the algorithm listed in the first column. The last column defines the used hyperparameters of all considered algorithms, all other unlisted hyperparameters were set as default for each respective algorithm implementation listed in the second column.}
\centering
\resizebox{\textwidth}{!}{
\begin{tabular}{cccc}
\hline
\textbf{Algorithm}	& \textbf{Version}	& \textbf{FIB Reference} & \textbf{Hyperparameters} \\ 
\hline
CB & Catboost 0.24.4 \citep{dorogush2018catboost} & - & Estimators: 550 \\
XGB & Xgboost 1.2.1  \citep{chen2016xgboost} & - & Estimators: 550 \\
RF & Scikit-learn 0.24.1  \citep{pedregosa2011scikit} & \citep{searcy2021day} & Estimators: 550 \\
SVR & Scikit-learn 0.24.1    & \citep{luvsic2017temporal} & Kernel: RBF / $\epsilon$: 0.23 / C: 20 \\
ANN & Scikit-learn 0.24.1   & \citep{searcy2021day} & Hidden layers: 2 / neurons: 100 \\
\hline
\label{tab:ml_algorithms}
\end{tabular}}
\end{table}

\section{Results and Discussion}
\subsection{ML Model Results}
\label{sub:ml_model}
In this section, an analysis of the accuracy of the ML model is assessed based on the coefficient of determination (R$^2$), and the root mean squared error (RMSE), a standard evaluation metric for regression. The K-fold cross validation model sampling method with k=10 and data shuffling was used to investigate the robustness of the model and the model was trained and tested with all of the routine monitoring data at all of the sampling locations except for location KS (a total of 1670 instances).

\begin{figure}
\makebox[\linewidth][c]{%
\centering
\subfloat{\includegraphics[width=0.7\textwidth]{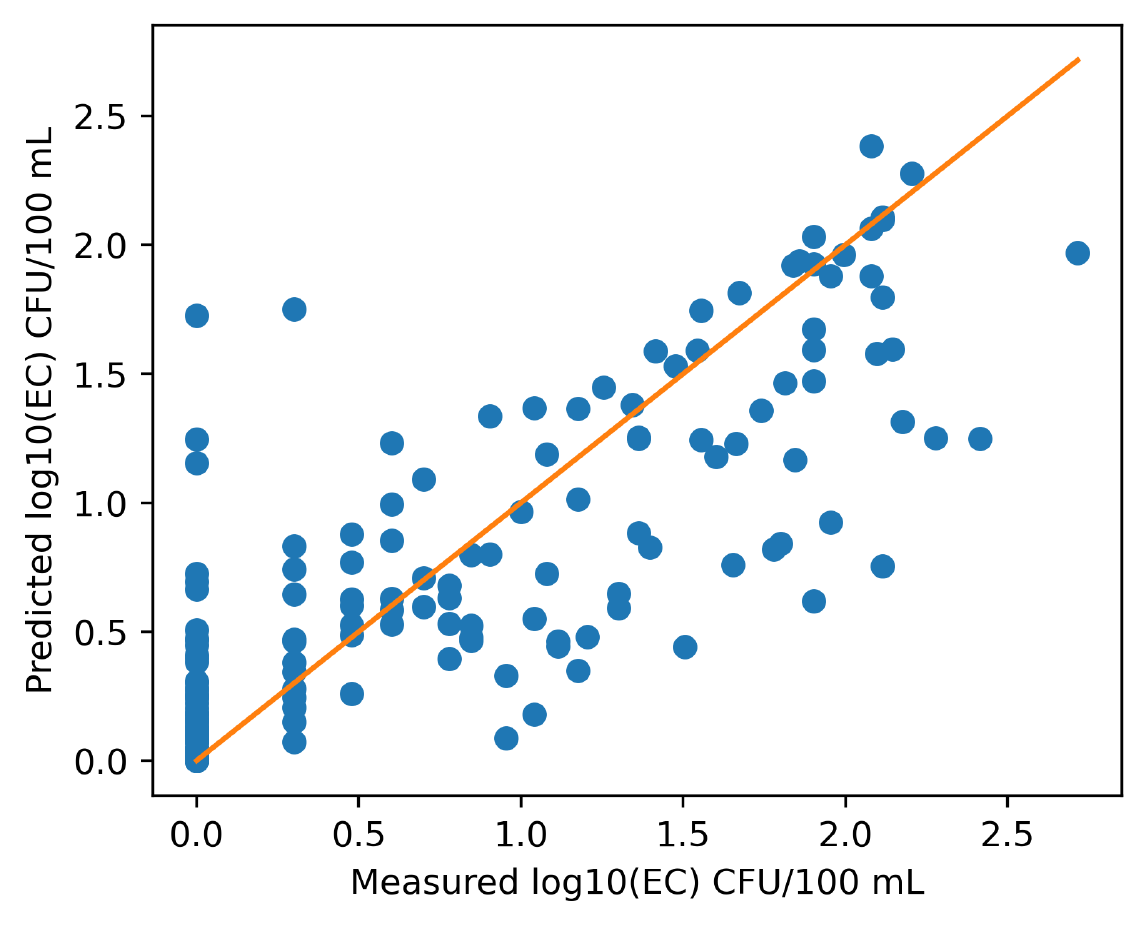}}%
\subfloat{\includegraphics[width=0.7\textwidth]{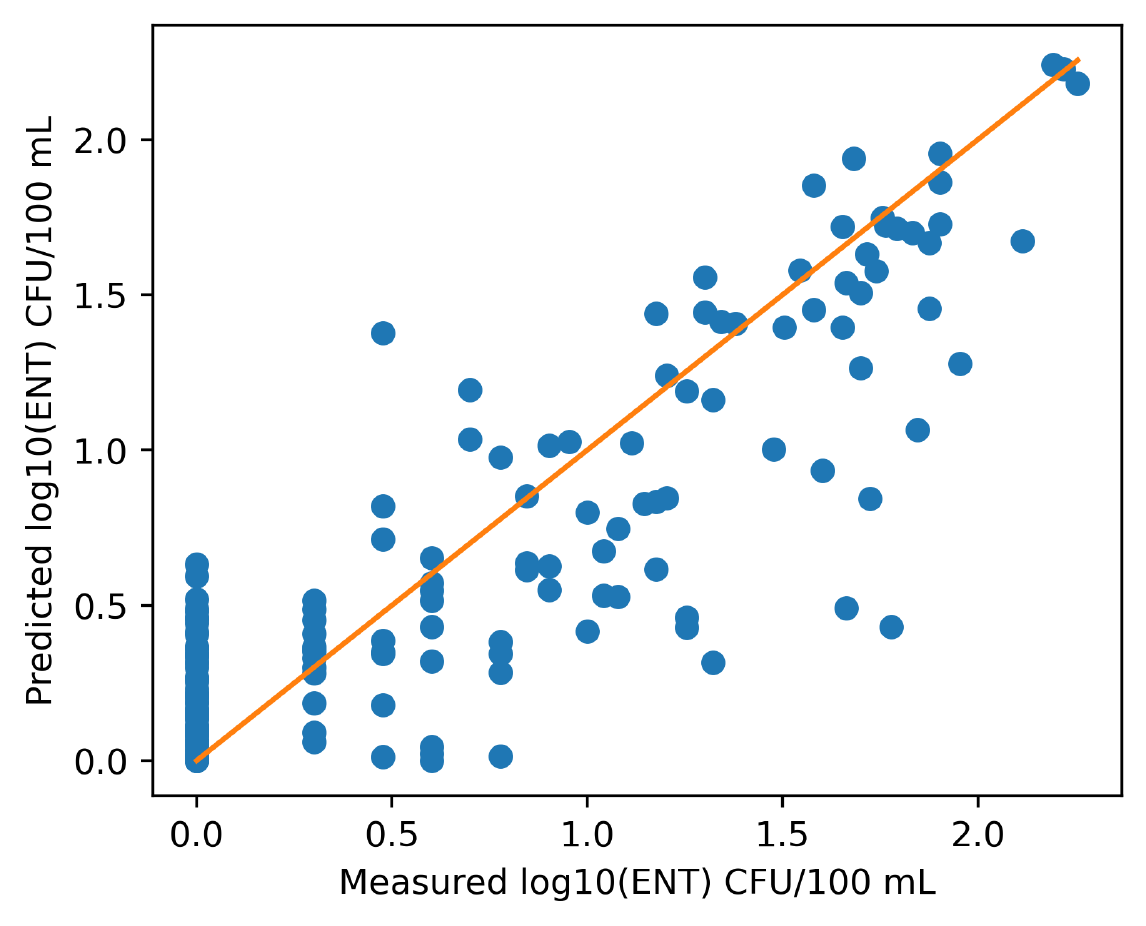}}%
}
\caption{The correlation between the measured and the CB model predicted values of both EC and ENT. The 10-fold cross validation generated a Spearman's rank correlation coefficient $\rho$ of 0.79 $\pm$ 0.03 for the EC model, and a $\rho$ of 0.77 $\pm$ 0.03 for the ENT model.}
\label{fig:correlation_graphs}
\end{figure}

Tables \ref{tab:model_results_ec} and \ref{tab:model_results_ent} present the EC and ENT models accuracy, respectively, and it can be seen that the most accurate algorithm in terms of both R$^2$ and RMSE is CB and that it also showed robustness in its K-fold validation since the standard deviation values are the low for both EC and ENT models. The RF algorithm is a close second in terms of accuracy metrics, while the SVR prediction can be considered the least accurate with a moderate R$^2$ score. Fig. \ref{fig:correlation_graphs} shows the correlation graphs of the CB trained ML models, and it can be observed that the models overestimate the measured FIB levels when they are zero.

\newpage

The ENT model RMSE values are overall lower than those obtained by the EC model, however, the R$^2$ metric is more useful in FIB predictive modeling than RMSE as it  shows the level of correlation between the input features and the output variable.

\begin{table}[!h]
\caption{The K-fold cross validation EC predictive model values of R$^2$ and RMSE with a standard deviation (std) for all of the investigated ML algorithms. The values which are in bold represent the most accurate result in terms of the given metric.}
\centering
\resizebox{\textwidth}{!}{
\begin{tabular}{cccccc}
\hline
\textbf{Metric}	& \textbf{RF}	& \textbf{XGB} & \textbf{CB} & \textbf{SVR} & \textbf{ANN}\\
\hline
R$^2$ $\pm$ std & 0.70 $\pm$ 0.04 & 0.66 $\pm$ 0.07 & \textbf{0.71 $\pm$ 0.06} & 0.63 $\pm$ 0.05 & 0.64 $\pm$ 0.06 \\
RMSE $\pm$ std & 0.43 $\pm$ 0.03 & 0.45 $\pm$ 0.03 & \textbf{0.43 $\pm$ 0.03} & 0.48 $\pm$ 0.05 & 0.46 $\pm$ 0.04 \\
\hline
\label{tab:model_results_ec}
\end{tabular}}
\end{table}

\begin{table}[!h]
\caption{The K-fold cross validation ENT predictive model values of R$^2$ and RMSE with a standard deviation (std) for all of the investigated ML algorithms. The values which are in bold represent the most accurate result in terms of the given metric.}
\centering
\resizebox{\textwidth}{!}{
\begin{tabular}{cccccc}
\hline
\textbf{Metric}	& \textbf{RF}	& \textbf{XGB} & \textbf{CB} & \textbf{SVR} & \textbf{ANN}\\
\hline
R$^2$ $\pm$ std & 0.66 $\pm$ 0.07 & 0.64 $\pm$ 0.06 & \textbf{0.68 $\pm$ 0.05}  & 0.61 $\pm$ 0.05 & 0.60 $\pm$ 0.04 \\
RMSE $\pm$ std & 0.38 $\pm$ 0.02 & 0.40 $\pm$ 0.03 & \textbf{0.38 $\pm$ 0.03} & 0.42 $\pm$ 0.03 & 0.41 $\pm$ 0.04 \\
\hline
\label{tab:model_results_ent}
\end{tabular}}
\end{table}
\newpage

\subsection{ML Model Feature Interpretation}
In this section, a feature analysis is presented using the game theoretic-based method SHapley Additive exPlanations (SHAP) \citep{NIPS2017_7062}. Generally, the SHAP values generated by the method indicate a feature's control over a change in the model output and it is used for ML model interpretation. 

SHAP is a powerful tool that has been recently used in various cutting edge research areas such as explaining the ML prediction of hypoxaemia prevention during surgery \citep{lundberg2018explainable}, interpreting the ML model-based behavior of nanophotonic structures \citep{yeung2020elucidating}, and explaining the relationship between the stream water quality (which includes EC measurements) with urban development patterns \citep{wang2021predicting}.

The SHAP approach is used to interpret the output of any ML model, but in this specific study it is used to interpret how the input features are related to the output values of the most accurate - CB model with a 90/10 training/testing data split. The feature with the highest mean SHAP value contributes the most to the model output (which are the EC and ENT values in this study), and therefore has the highest predictive power. 

The Tree Explainer algorithm \citep{lundberg2020local2global} was used which is implemented in the Shap 0.38.1. module for Python 3.8 and works in conjunction with decision tree ensemble based ML algorithms, specifically, in this case the CB algorithm. 

\subsubsection{EC Model Features}

\begin{figure}[!h]
\centering
\includegraphics[width=\textwidth]{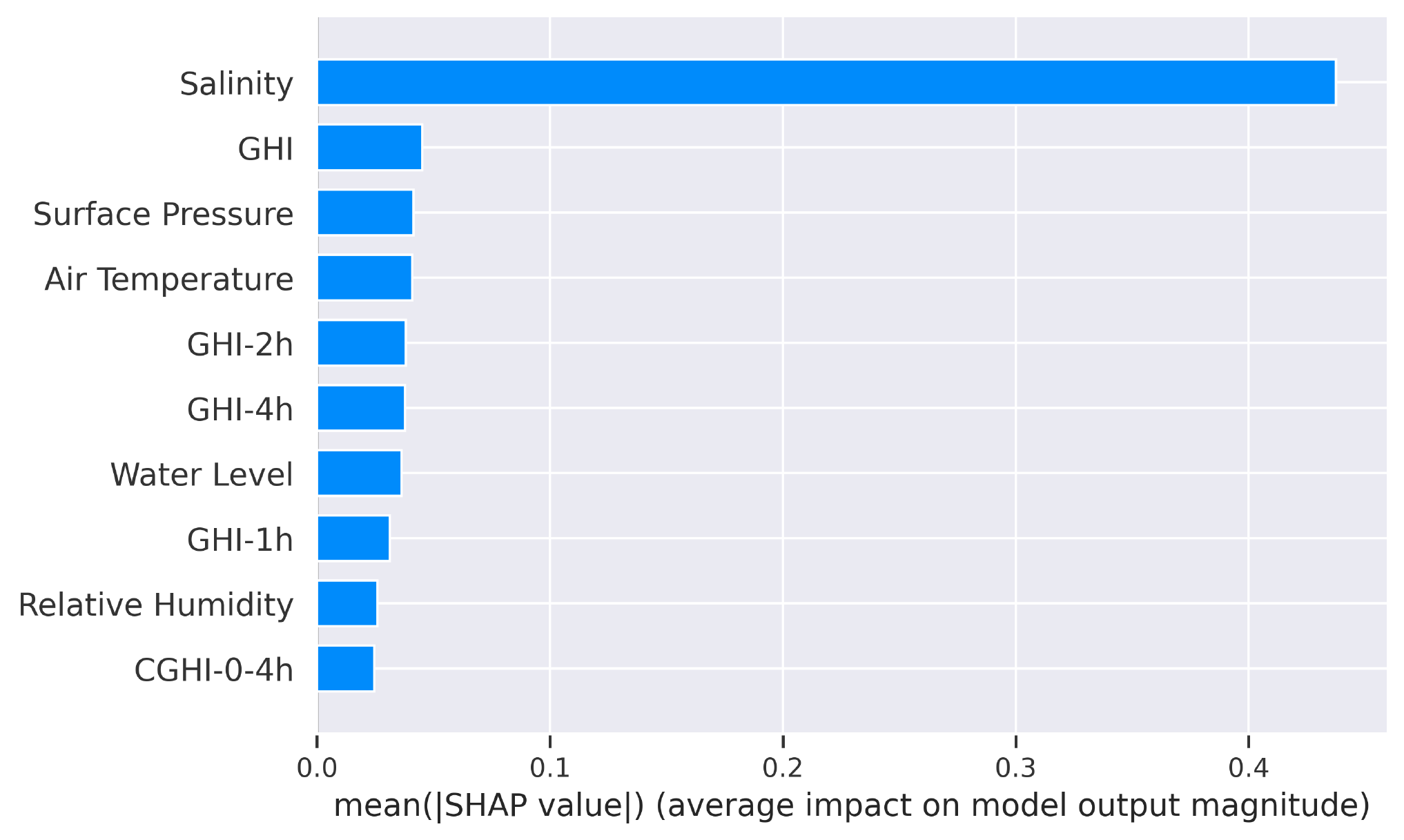}
\caption{SHAP feature importance values for the EC model. Feature with the highest SHAP value has the greatest impact on the EC prediction.}
\label{fig:shap_ec}
\end{figure}

Feature importance of the ten most significant features from the EC model, investigated by the SHAP method, is presented in a descending order in Fig. \ref{fig:shap_ec}. 

In this case, the most significant feature by far is the seawater salinity.
An interpretation of the observed result is that salinity has the power to change the predicted value of EC on average by above 40\%. All of the other considered features have an impact on the predicted value of EC, albeit to a lesser extent than salinity. The effects of salinity on EC and FIB in general have been studied in previous literature where it was recognized as an important environmental stressor and a catalyst for decay \citep{troussellier1998responses,jozic2014effect}.

It can be observed that the GHI value at EC measurement time and three other of the considered antecedent GHI variations have an impact on the EC value. Other significant features include surface pressure, air temperature, water level, relative humidity at measurement time and the antecedent 4 hour cumulative GHI before the measurement time. Even though only the top 10 features are shown in Fig. \ref{fig:shap_ec}, none of the other considered features had a 0\% significance over the EC prediction.

\begin{figure}[!h]
\centering
\includegraphics[width=\textwidth]{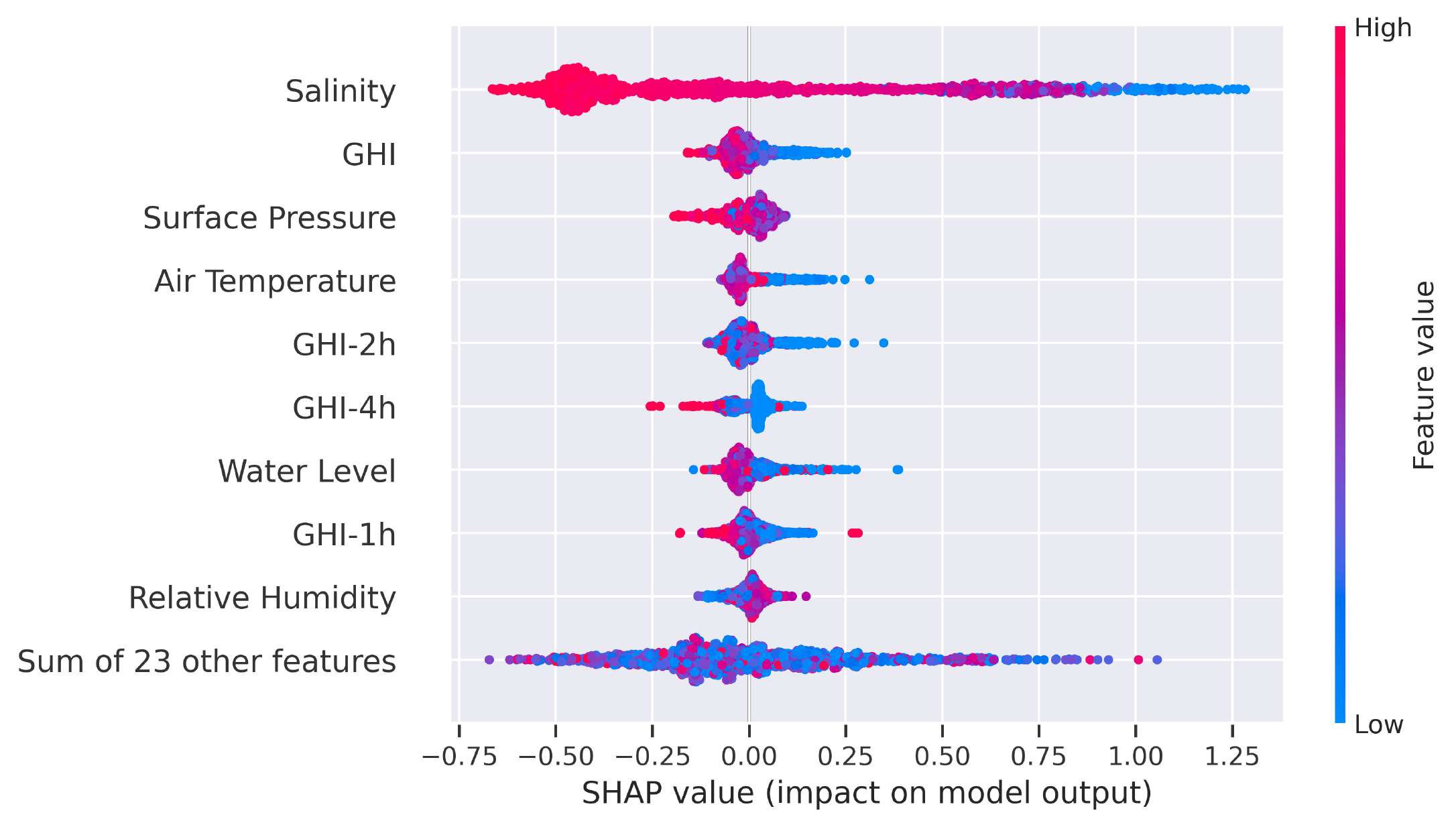}
\caption{SHAP feature impact values on EC prediction and correlations. Red and blue colors indicate high and low EC values, respectively.}
\label{fig:shap_ec_swarm}
\end{figure}

Fig. \ref{fig:shap_ec_swarm} shows SHAP values of each feature along with their respective positive and negative relationships with the EC prediction value. The features are also ranked by importance in descending order as in Fig. \ref{fig:shap_ec} and the x-axis represents the SHAP values along with the impacts of each feature on the EC model output and their correlation. The red color indicates if the value of EC is high, while the blue color represents the low value of EC. For example, when the salinity at EC measurement time is high (SHAP value above 1), the EC value is low, i.e. high salinity has a negative impact on EC. If the GHI value is low (SHAP value below 0), the EC prediction is expected to be high, which indicates a negative correlation between the two features, while relative humidity exhibits a positive correlation with EC values. A negative correlation between GHI and EC is expected and in line with previous laboratory and environmental FIB (which includes EC and ENT) sampling and modeling research \citep{jozic2014effect,davies1994sunlight,alkan1995survival}. The sum of the 23 remaining features shows a wide and complex impact on the EC values, additionally showing the sensitivity and non-linearity of the studied problem.

\begin{figure}[!h]
\centering
\includegraphics[width=\textwidth]{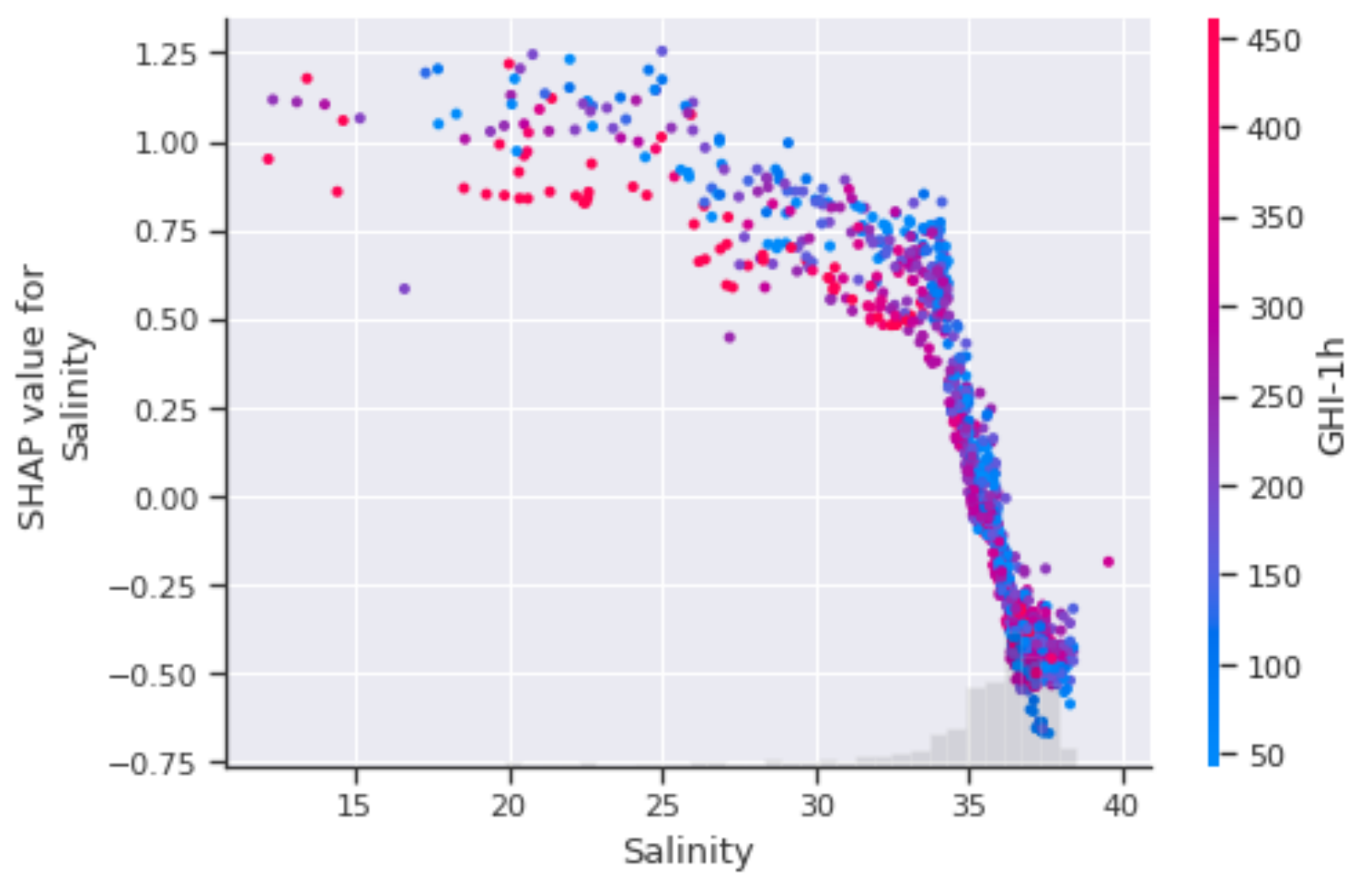}
\caption{Relationship between salinity and the EC prediction SHAP values. Color indicates relationship between salinity and 1 hour antecedent GHI.}
\label{fig:salinity_EC}
\end{figure}

Fig. \ref{fig:salinity_EC} shows the relationship between the salinity values (x-axis) and EC SHAP values (y-axis), i.e. how the model prediction of EC changes with the increase or decrease of salinity. An interesting phenomena can be observed as there exists a negatively linear correlation between the predicted change of EC and salinity when the value of salinity is approximately above 34.

The area around the easternmost cluster of the sampling locations which includes KE, KW and KS has a direct groundwater connection with an existing reservoir and a river as it was shown in previous research \citep{biondic1997hydrogeological,bonacci2018water}. A strong relationship between salinity and predicted EC values by the model could indicate that the studied locations feature active fresh water springs - end points of a complex groundwater network - which are the main source of EC inflow into the marine area.

Additionally, SHAP analysis indicates that the GHI measured 1 hour before the EC measurement time closely interacts with salinity in a way that an approximately higher 1 hour antecedent GHI value corresponds to a decreased salinity.

\newpage

\subsubsection{ENT Model Features}

\begin{figure}[!h]
\centering
\includegraphics[width=\textwidth]{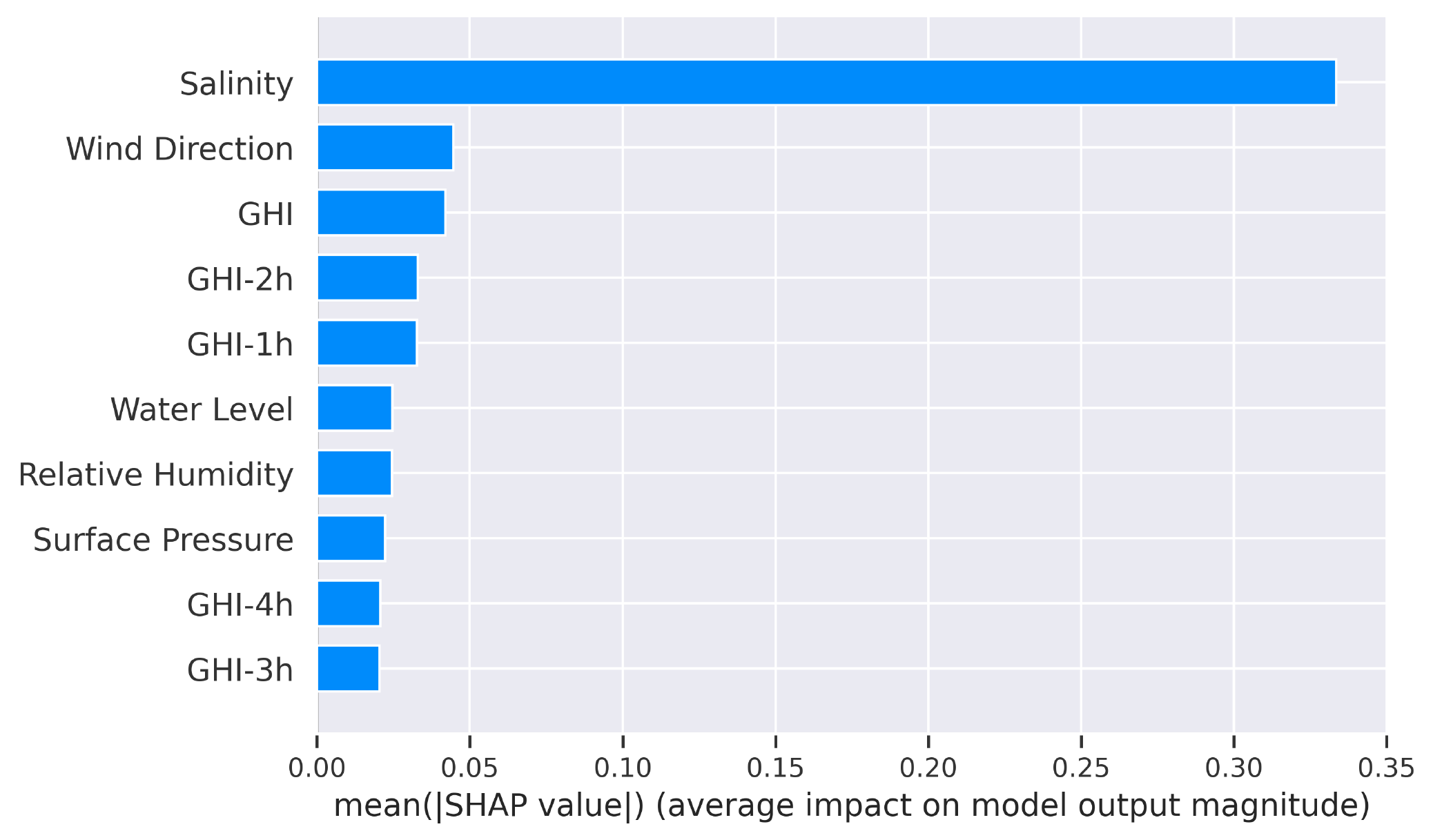}
\caption{SHAP feature importance values for the ENT model.}
\label{fig:shap_ent}
\end{figure}

SHAP analysis has also been performed for the ENT prediction model, and in Fig. \ref{fig:shap_ent} the SHAP feature importance values can be observed. Just as for the EC model, salinity remained the dominant feature that influences the output values of the ENT model. GHI and all of the antecedent GHI features continue to have an impact on ENT as well, while surface pressure, water level and relative humidity also show an influence on the model output.

Interestingly, the wind direction is a feature which has an effect on the ENT model and its SHAP value makes it the second most important feature that controls the output prediction as the features are sorted in a descending order by their importance.

Furthermore, in Fig. \ref{fig:shap_ent}, the negative and positive correlations between the features and the predicted ENT values can be seen. Of the most impactful features, both salinity and GHI have a negative correlation with the ENT predictions, while wind direction has a positive relationship with ENT as interpreted by SHAP. A similar observation can be derived for the ENT model as the one for EC, and it is that the major source of ENT could be the fresh water entering the coastal area through the karst terrain and crevices which characterize the studied cluster. In a review given by \citep{byappanahalli2012enterococci} both salinity and sunlight have been proven to have a negative effect on ENT values.

\begin{figure}[!h]
\centering
\includegraphics[width=\textwidth]{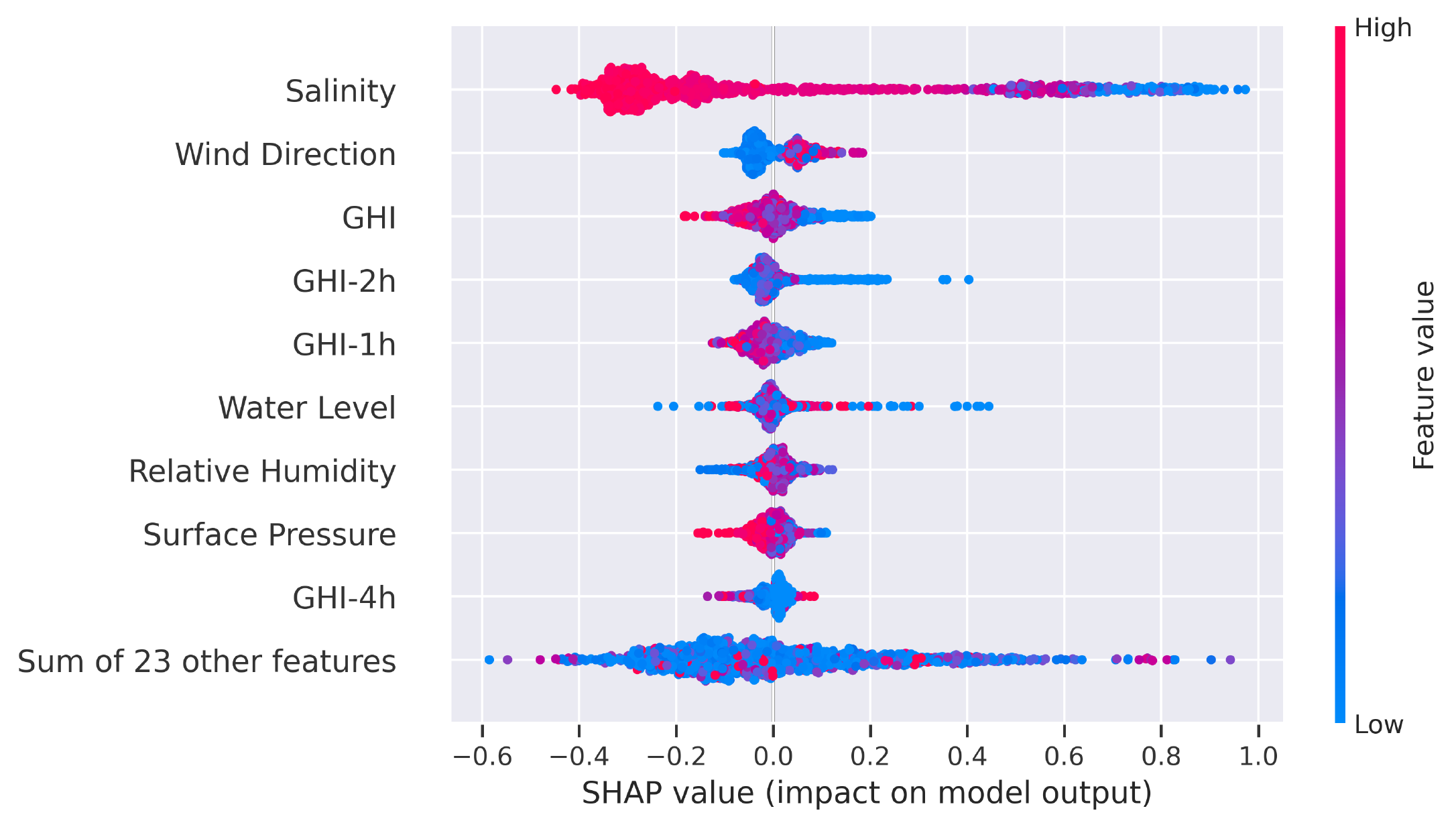}
\caption{SHAP feature impact values on ENT prediction and correlations. Red and blue colors indicate high and low ENT values, respectively.}
\label{fig:shap_ent_swarm}
\end{figure} 

\begin{figure}[!h]
\centering
\includegraphics[width=\textwidth]{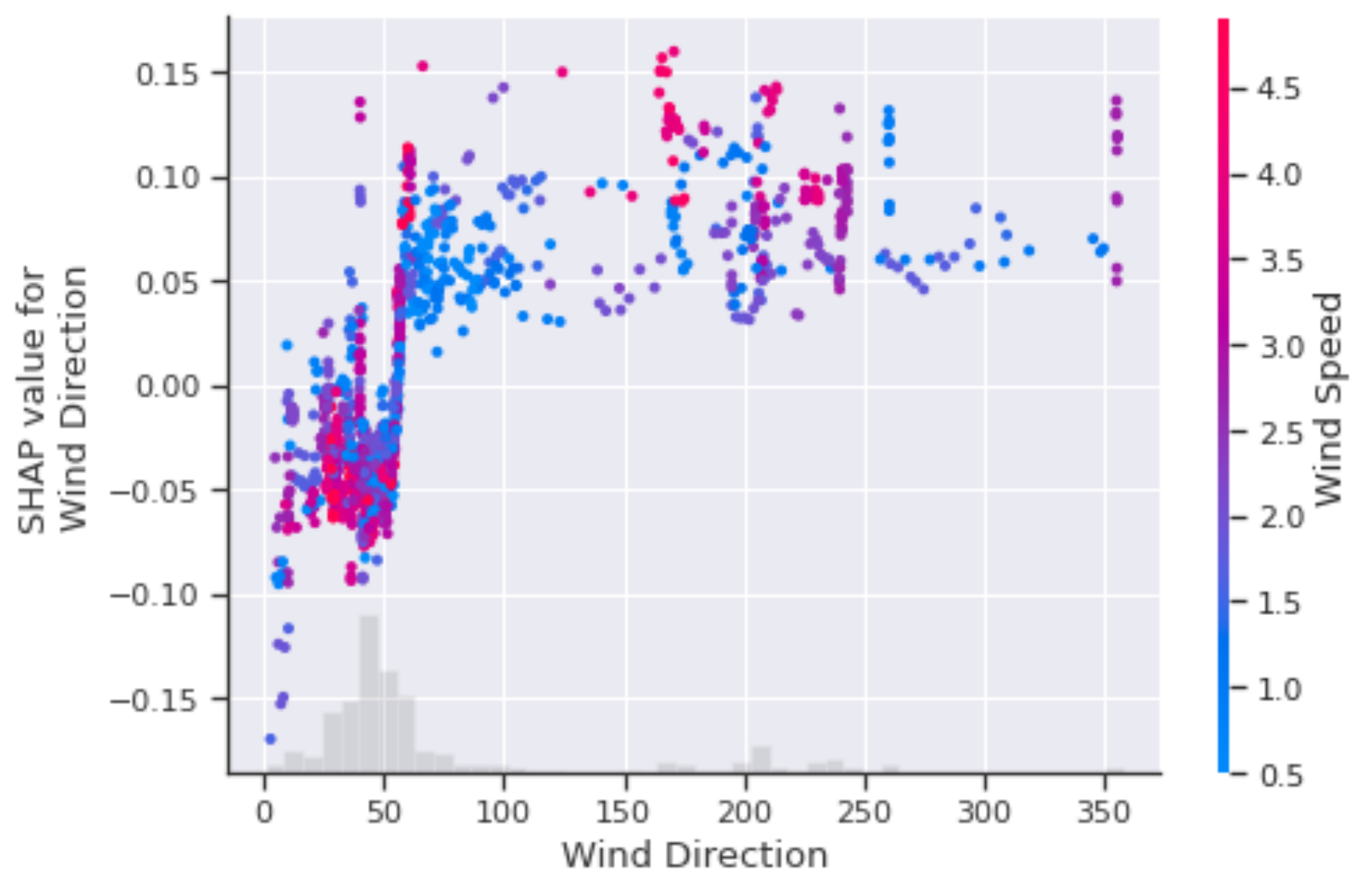}
\caption{Relationship between wind direction and the ENT prediction SHAP values. Color indicates relationship between wind direction and wind speed. The north-eastern wind (\textit{Bora}) has generally a greater wind speed than other winds typical for the area.}
\label{fig:wind_dir}
\end{figure}

Since the wind direction feature has an impact on the value of ENT, it could be hypothesized that ENT is more likely to be transported via sea currents than EC. This phenomena was also observed in literature \citep{liu2006modeling}, where a wind driven circulation hydrodynamic model was coupled with a bacterial transport model which incorporates a decay rate, and when compared with in situ measurements, it was found that ENT has a higher persistence than EC. This can also be observed in Fig. \ref{fig:median_ec_ent}, even though the median values of EC are higher, especially at the eastern part of the studied cluster, the presence of ENT is more observable at westernmost sampling locations. Finally, in Fig. \ref{fig:wind_dir}, a detailed relationship of wind direction and the model output can be seen, where a wind direction of around 45 degrees would be characteristic of the wind blowing from the north-east of the studied area, while a 200 degree wind direction indicates a south-western wind.

\newpage
\subsection{Spatial ML Prediction}

In this subsection a spatial prediction ML model is investigated. The spatial ML model is used to predict FIB values with input feature data from a location which is not used for model training.

Accurate spatial prediction of EC or ENT could be useful for determining points where the coastal water quality is low, and that could indicate a presence of an underground source of fresh water which is then transported to other locations with surface currents, therefore ultimately polluting a larger area. Additionally, an accurate spatial model could be useful for reducing the number of FIB sampling points.

To examine ML spatial prediction, a model was trained with routine monitoring data collected at all sample locations except for KS, which only includes measurements from 2019 to 2020. KS measurements were used as test data to assess the accuracy of the model, and it is a problematic location in terms of water quality as it is located right next to the easternmost part of the cluster (KE and KW) which historically has the highest average and median values of both EC and ENT.

All of the algorithms used to train and test the general ML model in section \ref{sub:ml_model} were used to train and test the spatial ML model. Number of training data instances were 1670, while the KS test data had 20 days, a total of 10 for 2019 and 10 for 2020. All of the considered features were used to train and test the spatial ML model.

In Tables \ref{tab:spatial_ec} and \ref{tab:spatial_ent}, the results of the EC and ENT spatial ML models are presented, respectively. Similarly as for the general model ML model, the model trained with the Catboost algorithm outperforms the other considered models, and the EC model is more accurate for both considered metrics than the ENT model.

\newpage

\begin{table}[!h]
\caption{The EC spatial predictive model values of R$^2$ and RMSE for all of the investigated ML algorithms. The values which are in bold represent the most accurate result in terms of the given metric.}
\centering
\begin{tabular}{cccccc}
\hline
\textbf{Metric}	& \textbf{RF}	& \textbf{XGB} & \textbf{CB} & \textbf{SVR} & \textbf{ANN}\\
\hline
R$^2$ & 0.82 & 0.77 & \textbf{0.85} & 0.41 & 0.24  \\
RMSE & 0.37 & 0.42 & \textbf{0.34} & 0.66  & 0.76 \\
\hline
\label{tab:spatial_ec}
\end{tabular}
\end{table}

\begin{table}[!h]
\caption{The ENT spatial predictive model values of R$^2$ and RMSE for all of the investigated ML algorithms. The values which are in bold represent the most accurate result in terms of the given metric.}
\centering
\begin{tabular}{cccccc}
\hline
\textbf{Metric}	& \textbf{RF}	& \textbf{XGB} & \textbf{CB} & \textbf{SVR} & \textbf{ANN}\\
\hline
R$^2$ & 0.75  & 0.79 & \textbf{0.83}  & 0.39 & 0.38  \\
RMSE & 0.39  & 0.35 & \textbf{0.31} & 0.60 & 0.60 \\
\hline
\label{tab:spatial_ent}
\end{tabular}
\end{table}

Fig. \ref{fig:ec_ent_spatial} shows the EC (Fig. \ref{fig:ec_spatial}) and ENT (Fig. \ref{fig:ent_spatial}) spatial prediction for the KS location with the CB algorithm. Primarily, a high R$^2$ for the spatial prediction of KS indicates that this location has very similar EC and ENT dynamics as its neighboring locations KE and KW. Furthermore, this shows that the spatial ML model could be used to predict FIB values at a location which wasn't included in the model training data.

Additionally, it can be observed that the EC model is better at predicting the peak values of EC measurements than the ENT model, and that both EC and ENT measurements follow a similar trend in 2019 and 2020. 

\newpage

\begin{figure}
  \centering
  \subfloat[a][]{\includegraphics[width=0.85\textwidth]{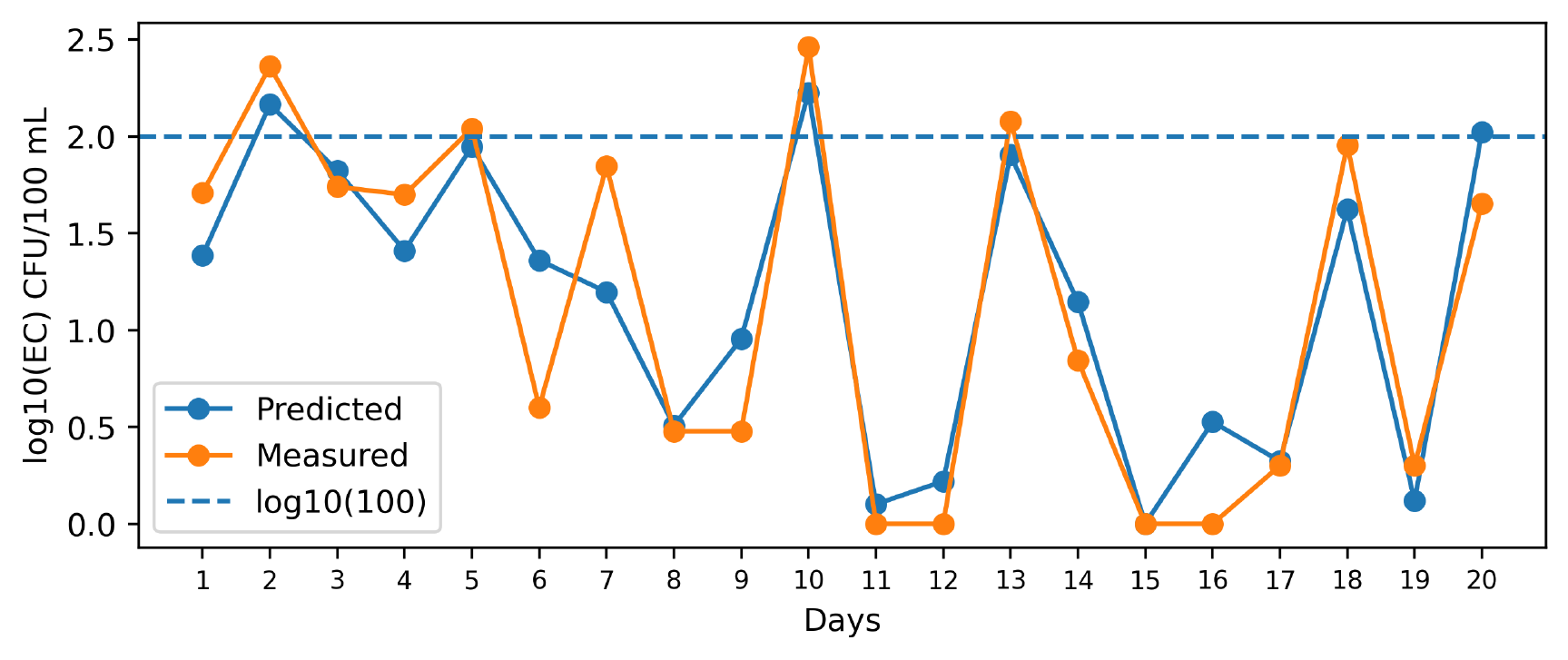} \label{fig:ec_spatial}} \\
  \subfloat[b][]{\includegraphics[width=0.85\textwidth]{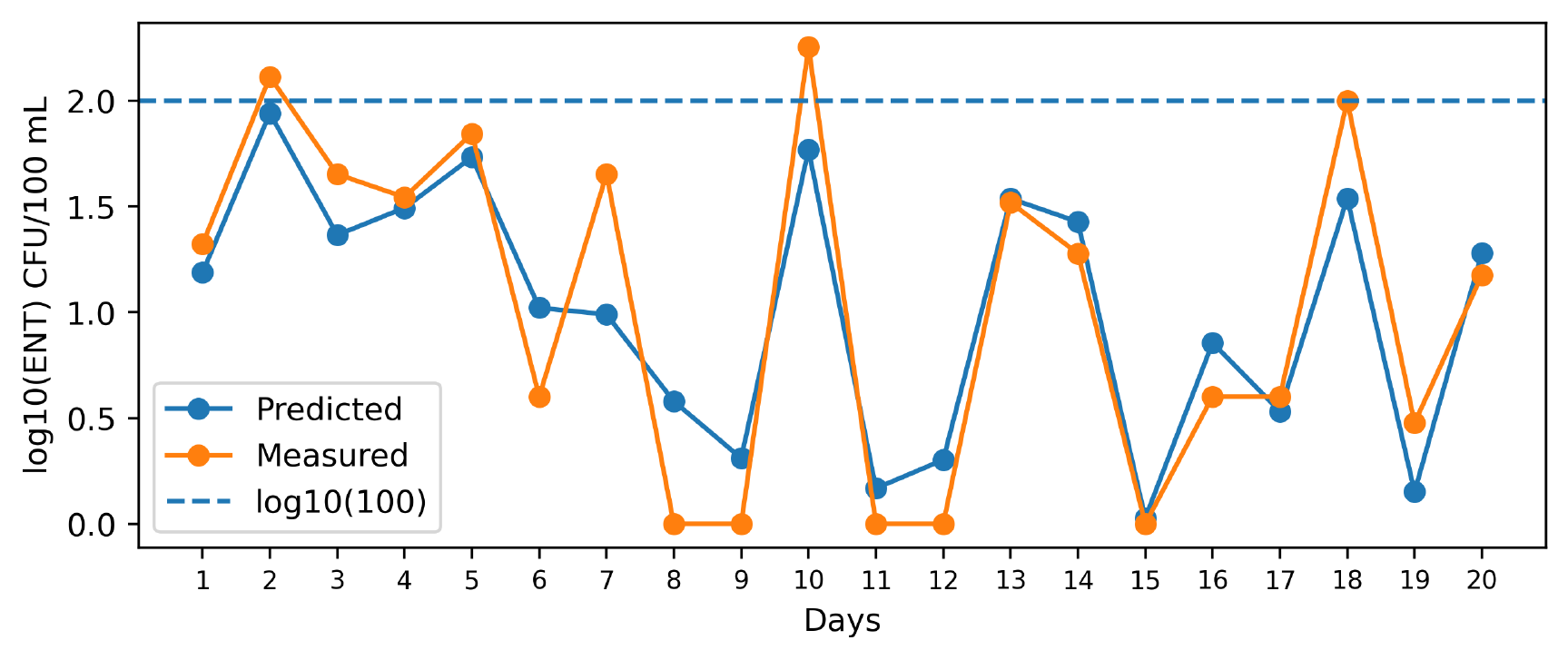} \label{fig:ent_spatial}}
  \caption{EC and ENT spatial ML prediction for KS (2019-2020) with the CB algorithm. The EC model has a R$^2$ value of 0.85, while the ENT model has 0.83. The 20 days are in chronological order.} 
  \label{fig:ec_ent_spatial}
\end{figure}

\subsection{Temporal ML Prediction}

An accurate temporal ML prediction is extremely beneficial since it would allow for public health authorities to issue a warning in advance if a certain location is safe for bathing or not.

\subsubsection{KE, KW and KS Temporal Prediction for 2020}

Firstly, a ML model was trained with routine monitoring data which was collected before 2020 at all locations (except KS) in order to predict the EC and ENT values for the year 2020 at KE, KW and KS. An assessment of KE, KW and KS is extremely useful as they are the three most problematic locations with the poorest quality of coastal water. Additionally, the distance between KE and KW is 100 meters, while KW and KS have a 250 meter gap between them.

All considered features were used to train the model and all previously described ML algorithms were investigated. The number of instances for model training was 1550, while 30 measurements in 2020 were used for model testing, where KE, KW and KS included 10 measurements at each location.

The results are presented in Tables \ref{tab:temporal_ec_kekwks} and \ref{tab:temporal_ent_kekwks}. In the case of 2020 temporal prediction for KE, KW and KS, the EC model achieved a greater accuracy than the ENT model since the results produced have a stronger R$^2$ score, while for the ENT model the R$^2$. The RF algorithm achieved the second best score for both EC and ENT models, while the negative values for both SVR and ANN indicate that a horizontal line would produce a better agreement with the data than the generated values.

\begin{table}[!h]
\caption{The EC temporal predictive model values of R$^2$ and RMSE for all of the investigated ML algorithms for the KE, KW and KS 2020 measurements. The values which are in bold represent the most accurate result in terms of the given metric.}
\centering
\begin{tabular}{cccccc}
\hline
\textbf{Metric}	& \textbf{RF}	& \textbf{XGB} & \textbf{CB} & \textbf{SVR} & \textbf{ANN}\\
\hline
R$^2$ & 0.69 & 0.56 & \textbf{0.74} & -0.25 & -0.07  \\
RMSE & 0.42 & 0.5 & \textbf{0.39} & 0.85  & 0.78 \\
\hline
\label{tab:temporal_ec_kekwks}
\end{tabular}
\end{table}

\newpage

\begin{table}[!h]
\caption{The ENT temporal predictive model values of R$^2$ and RMSE for all of the investigated ML algorithms for the KE, KW and KS 2020 measurements. The values which are in bold represent the most accurate result in terms of the given metric.}
\centering
\begin{tabular}{cccccc}
\hline
\textbf{Metric}	& \textbf{RF}	& \textbf{XGB} & \textbf{CB} & \textbf{SVR} & \textbf{ANN}\\
\hline
R$^2$ & 0.64  & 0.51 & \textbf{0.67}  & -0.74 & -1 \\
RMSE & 0.33  & 0.39 & \textbf{0.32} & 0.74 & 0.79 \\
\hline
\label{tab:temporal_ent_kekwks}
\end{tabular}
\end{table}

In Fig. \ref{fig:ec_ent_spatial} the EC and ENT CB algorithm prediction graph at KE, KW and KS locations for 2020 is shown. The measured ENT and EC trends are similar at both locations, although there is only a single instance at location KS when the ENT value was over 100 CFU/100 mL.

\newpage

\begin{figure}
  \centering
  \subfloat[a][]{\includegraphics[width=0.85\textwidth]{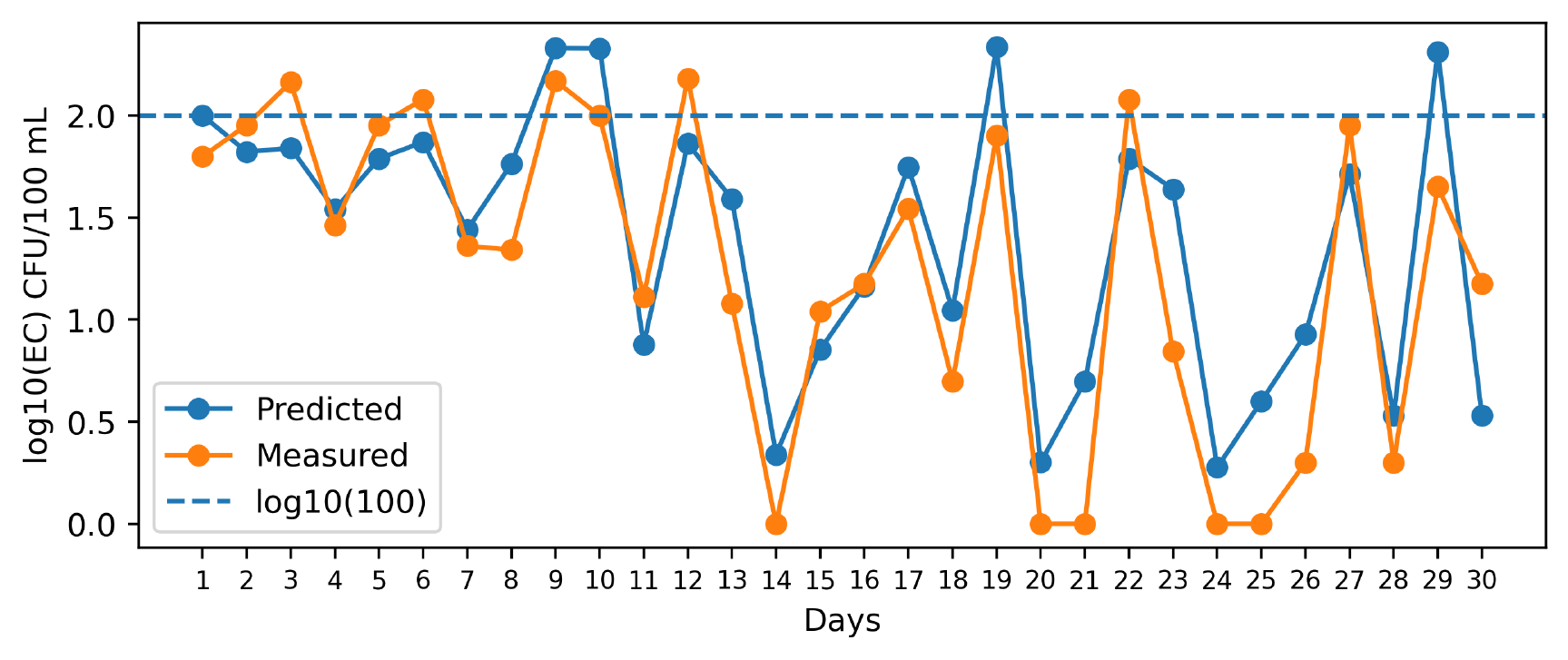} \label{fig:ec_spatial}} \\
  \subfloat[b][]{\includegraphics[width=0.85\textwidth]{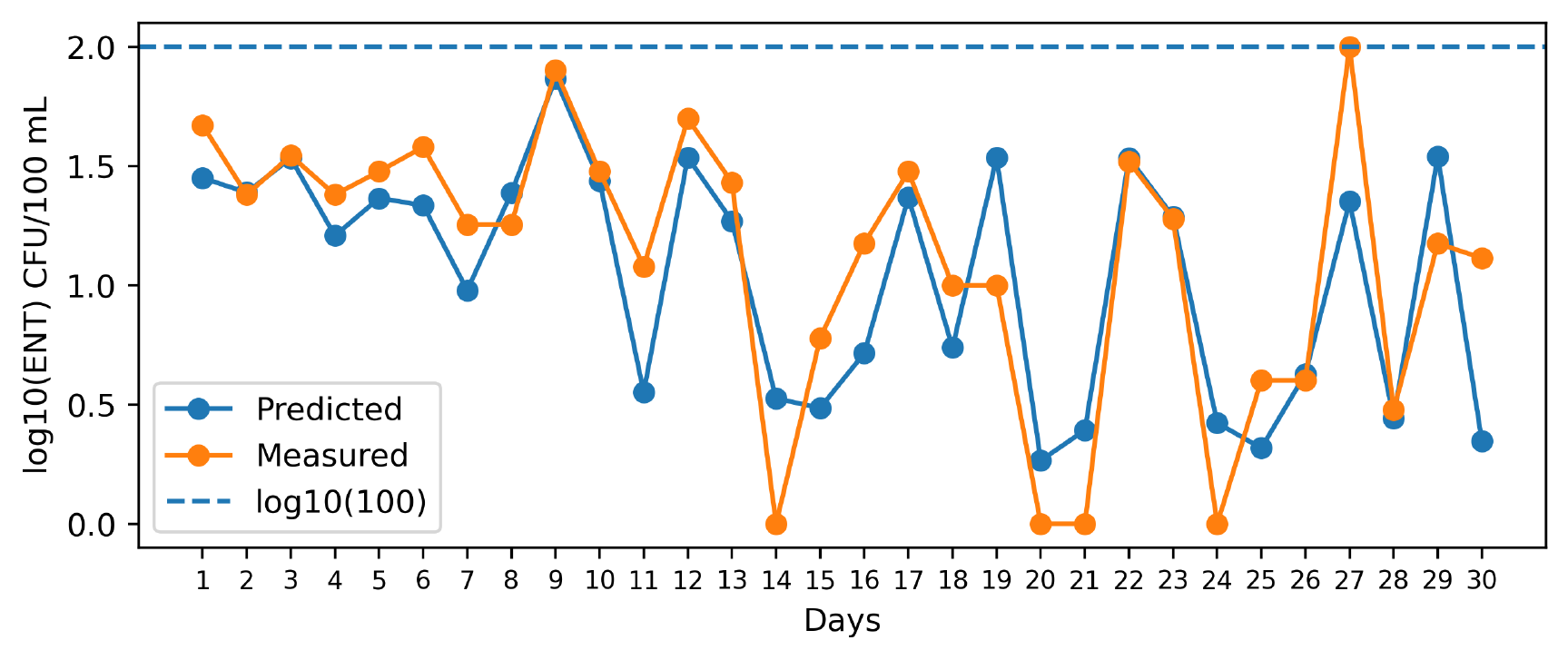} \label{fig:ent_spatial}}
  \caption{EC and ENT temporal ML prediction for KE, KW and KS (2020) with the CB algorithm. The EC model has a R$^2$ value of 0.74, while the ENT model has 0.67. The first 10 days are the KE measurements in chronological order, the next 10 days represent (11-20) KW measurements, and the last 10 days (21-30) represent location KS.} 
  \label{fig:ec_ent_spatial}
\end{figure}

\subsubsection{PW Temporal Prediction for 2020}

Lastly, the temporal prediction of the ML model trained with routine monitoring data collected before 2020 at all locations were tested on the data collected in 2020 at the westernmost point PW of the studied cluster. Even though PW is historically a location with high quality of sea water, it is useful to investigate how the temporal model predicts the EC and ENT values in an environment which is safe for coastal.

Similarly as for the KE, KW and KS model, all considered features were used to train the model and all previously used algorithms were assessed. The number of instances for model training was 1550, while 10 measurements at location PW in 2020 were used for model testing. 

\begin{figure}
  \centering
  \subfloat[a][]{\includegraphics[width=0.85\textwidth]{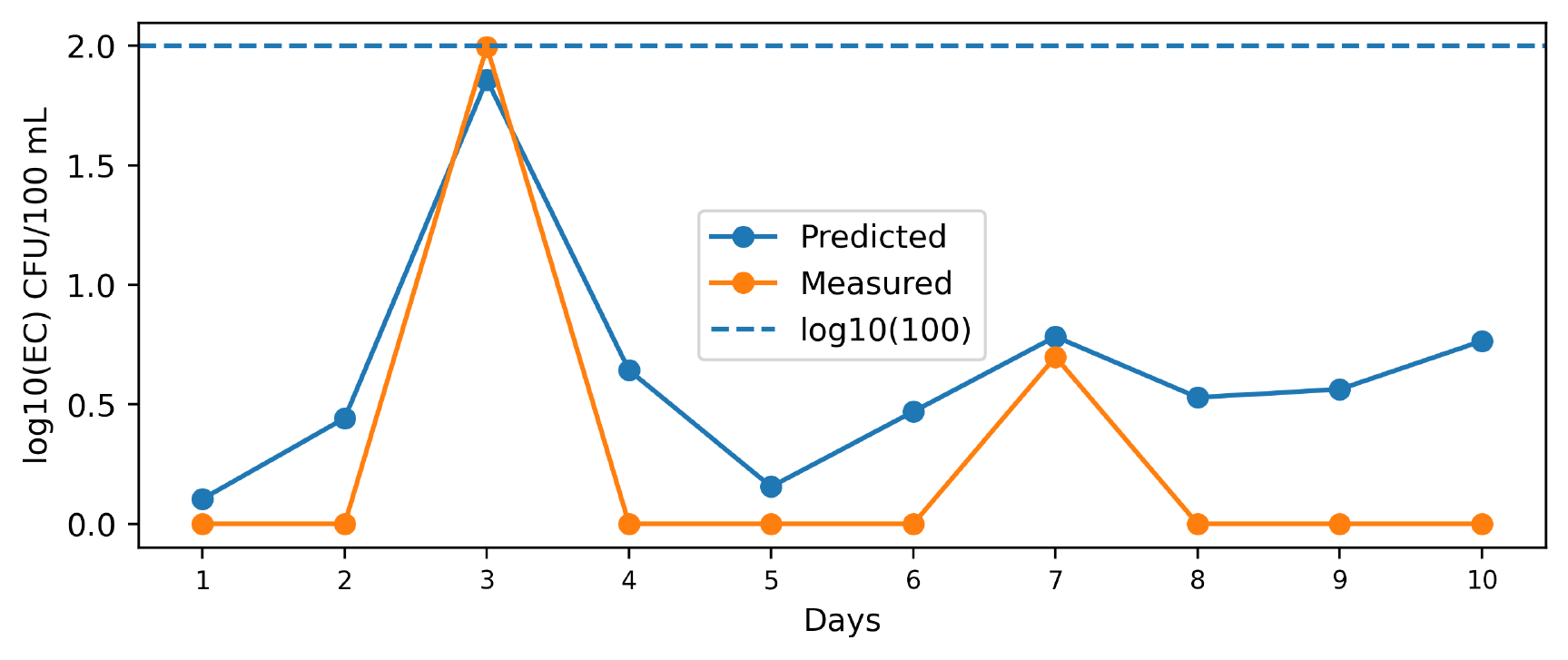} \label{fig:ec_temp_pw}} \\
  \subfloat[b][]{\includegraphics[width=0.85\textwidth]{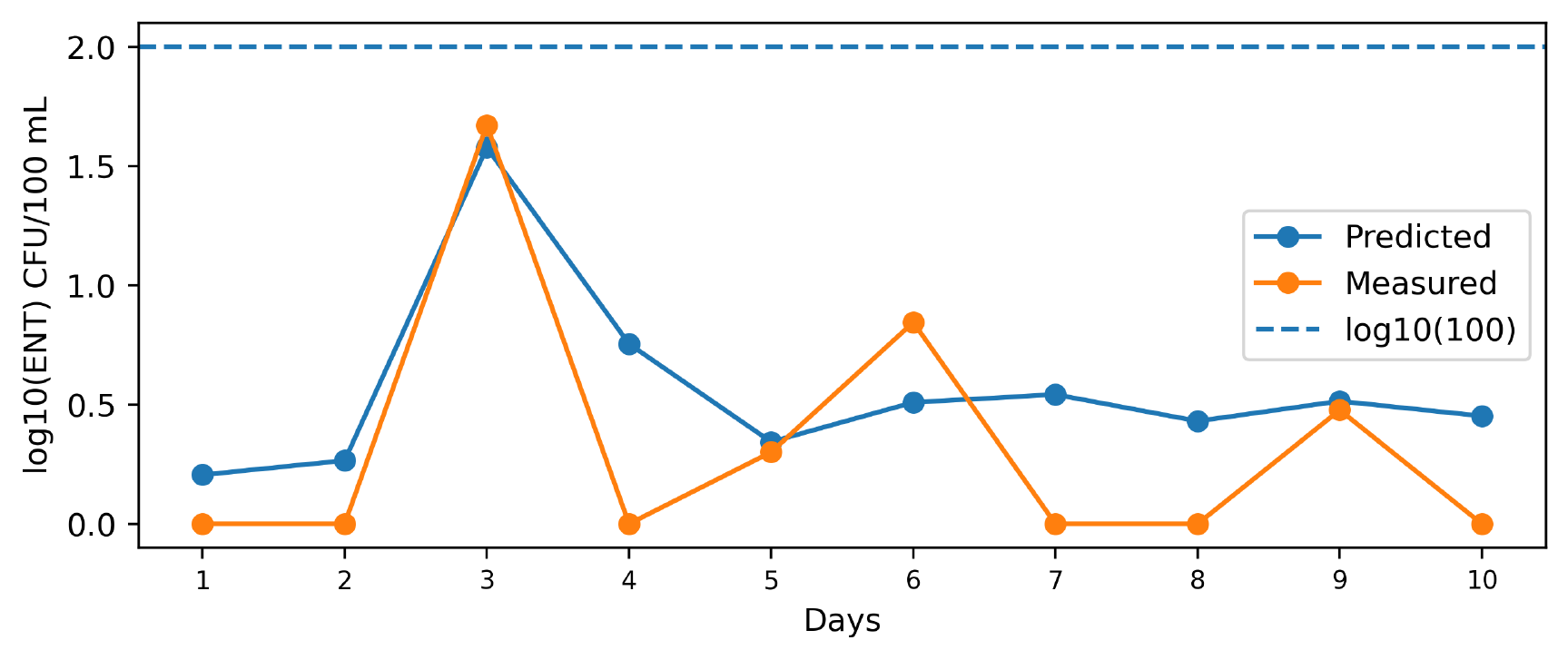} \label{fig:ent_temp_pw}}
  \caption{EC and ENT temporal ML prediction for PW (2020) with the CB algorithm. The EC model has a R$^2$ value of 0.44, while the ENT model has 0.46. The 10 days are in chronological order.} 
  \label{fig:ec_ent_temporal}
\end{figure}

The results for EC and ENT are presented in Tables \ref{tab:temporal_ec_pw} and \ref{tab:temporal_ent_pw}, respectively. The R$^2$ scores for both models are lower than those for the easternmost cluster of sampling locations, and similarly as the previous spatial and temporal models, the EC prediction is better than ENT. Even though the R$^2$ score indicates a weak to moderate predictive power of the model for the PW location, in Fig. \ref{fig:ec_ent_temporal} it can be observed that both EC and ENT models predict the maximum value with great accuracy, which is beneficial for this kind of problem. However, both models clearly overestimate the EC and ENT values when they were measured as 0, thus greatly reducing the R$^2$ score.

\newpage
\begin{table}[!h]
\caption{The EC temporal predictive model values of R$^2$ and RMSE for all of the investigated ML algorithms for the PW 2020 measurements. The values which are in bold represent the most accurate result in terms of the given metric.}
\centering
\begin{tabular}{cccccc}
\hline
\textbf{Metric}	& \textbf{RF}	& \textbf{XGB} & \textbf{CB} & \textbf{SVR} & \textbf{ANN}\\
\hline
R$^2$ & 0.42 & 0.22 & \textbf{0.44} & -0.46 & -1.32 \\
RMSE & 0.46 & 0.54 & \textbf{0.46} & 0.74  & 0.93 \\
\hline
\label{tab:temporal_ec_pw}
\end{tabular}
\end{table}

\begin{table}[!h]
\caption{The ENT temporal predictive model values of R$^2$ and RMSE for all of the investigated ML algorithms for the PW 2020 measurements. The values which are in bold represent the most accurate result in terms of the given metric.}
\centering
\begin{tabular}{cccccc}
\hline
\textbf{Metric}	& \textbf{RF}	& \textbf{XGB} & \textbf{CB} & \textbf{SVR} & \textbf{ANN}\\
\hline
R$^2$ & 0.32  & 0.18 & \textbf{0.46}  & -0.38 & -1.03  \\
RMSE & 0.46  & 0.47 & \textbf{0.38} & 0.61 & 0.75 \\
\hline
\label{tab:temporal_ent_pw}
\end{tabular}
\end{table}

\newpage

\newpage

\section{Conclusion}

In this research, EC and ENT routine monitoring data collected at 15 different sampling locations in the city of Rijeka, Croatia, were used to create spatial and temporal ML models for the purpose of predicting FIB values based on environmental features, and getting a better insight into the dynamics and correlations between FIB and environmental variables. The main contributions of this study can be summarized with the following points:

\begin{itemize}
    
    \item Firstly, both EC and ENT ML models were created with the routine monitoring data (colllected at all but one location) and environmental features. Gradient Boosting algorithms (Catboost, Xgboost), Random Forests, Artificial Neural Networks and Support Vector Regression were used to train and test the EC and ENT models. Cross validation showed that the Catboost algorithm performs the best with an achieved strong R$^2$ score of 0.71 for EC prediction and 0.68 for ENT. The Random Forest algorithm achieved the second best score for all trained EC and ENT ML models.
    
    \item Secondly, a spatial ML EC and ENT prediction was done on one location to investigate the accuracy of the model in that regard. 
    It was found that the model trained with the Catboost algorithm performs exceptionally well, achieving R$^2$ scores of 0.85 and 0.83 for EC and ENT predictions, respectively. A spatial model could show its usefulness in finding the locations of groundwater springs as they are severely linked with EC and ENT through salinity.

    \item Furthermore, a temporal ML EC and ENT prediction done in order to assess how well could the ML model be used to predict future EC and ENT values. The easternmost cluster which has historically been the worst in terms of water quality was investigated for the year 2020, and good R$^2$ scores of 0.74 and 0.67 were achieved. Additionally, the model achieved moderate to weak R$^2$ scores of 0.44 and 0.46 on the westernmost location with historically great quality of water.
    
    \item SHapley Additive exPlanations method was used to interpret which features are the most important in the ML model prediction. It was found that salinity is the most important feature for both EC and ENT model prediction. Other major contributors include the global horizontal irradiance for both EC and ENT, while wind direction also has an impact on the ENT prediction.
    
    \item Salinity as the most important feature could indicate that both EC and ENT are brought to the marine area by groundwater karst springs which are quite typical for the studied locations according to previous research. Furthermore, modeling the dynamics of salinity in conjunction with groundwater flow should be a topic of future research.

\end{itemize}

\section*{Acknowledgements}
This research article is a part of the project \textit{Computational fluid flow, flooding, and pollution propagation modeling in rivers and coastal marine waters--KLIMOD} (grant no. KK.05.1.1.02.0017), and is funded by the Ministry of Environment and Energy of the Republic of Croatia and the European structural and investment funds. Also, the authors acknowledge the funding of the University of Rijeka through the project \textit{Razvoj hibridnog 2D/3D modela za učinkovito modeliranje strujanja u rijekama, jezerima i morima}. Furthermore, authors acknowledge the support of the Center of Advanced Computing and Modelling at the University of Rijeka for providing computing resources for machine learning algorithm hyperparameter optimization.

\bibliographystyle{elsarticle-harv} 
\bibliography{cas-refs}

\end{document}